\documentclass[12pt]{article}
\pdfoutput=1
\usepackage{epsfig,graphicx,xcolor,amsbsy,amssymb,latexsym,amsfonts,amsmath,setspace}
\usepackage{pstricks}
\usepackage{color}
\usepackage{soul}
\usepackage[numbers,square,comma, compress]{natbib}
\usepackage{placeins}
\usepackage{wrapfig}
\usepackage{tikz}
\usepackage[a4paper]{geometry}
\usepackage{pst-node}
\usepackage{tikz-cd}
\usepackage{subfig}

\usepackage{graphicx}

\usepackage{xcolor}

\usepackage[enableskew]{youngtab}
  
\usepackage{etex}
\usepackage{relsize}
\usepackage{natbib}
\usepackage{amsfonts}
\usepackage[T1]{fontenc}
\usepackage{hyperref}

\usepackage{authblk}
\usepackage{blindtext}

\def\bea{\begin{eqnarray}}
\def\eea{\end{eqnarray}}

\usepackage{appendix}

\usetikzlibrary{decorations.markings,arrows,backgrounds,patterns,decorations.pathreplacing}

\begin{document}

\title{ 
 Comments on the little string partition functions of $K3\times T^2$ via the refined topological vertex
}

\author{Ambreen Ahmed\thanks{\href{mailto:ambreen.ahmed@uon.edu.pk}{ambreen.ahmed@uon.edu.pk}}
\and M. Nouman Muteeb\thanks{\href{mailto:nouman01uet@gmail.com}{nouman01uet@gmail.com}}}





\maketitle

\begin{abstract}
We compute partition functions of the deformed multiple M5-branes theory on $K3\times T^2$ using the refined topological vertex formalism and the Borcherds lift. The deformation is related to the mass deformation in the corresponding four dimensional $N=4$ $SU(N)$ gauge theory on $K3$. The seed of the 
 Borcherd-lift is calculated by taking the universal part of the type IIb little string free energy of the CY3-fold $X_{N,1}$. We provide explicit modular covariant expressions, as expansions in the mass parameter $m$, of the genus two Siegel modular forms  produced by  the Borcherds lift of the first few seed functions. 
 We also discuss the relation between genus-one free energy and Ray-Singer Torsion, and the automorphic properties of the latter.
\end{abstract}
\newpage
\tableofcontents
\newpage
\section{Introduction and summary }
Six dimensional SCFTs do not admit any obvious Lagrangian description in terms of UV degrees of freedom. Embedding these QFTs in M-theory/F-theory has proved to be very useful to study their dynamics. Following this line of inquiry a classification of of 6d SCFTs has been proposed by considering F-theory compactification on elliptically fibered Calabi-Yau  three folds (CY3-folds)
An exciting class of such theories are the so-called  Little string theories \cite{Dijkgraaf:2007sw,Dijkgraaf:2002ac,Dijkgraaf:1996it,Dijkgraaf:1996xw,Aharony:1999ks}, which are the descendent of usual strings in the limit  of vanishing string coupling constant  $g_s\to 0$, planck scale approaching infinity and finite string length  $l_s$. In type II string theory these strings are stuck to five-branes and do not give rise to spin 2  degrees of freedom.\\
Type  IIA little strings of $A_{N-1}$ type possess $N=(1,1)$ supersymmetry, whereas type IIB little strings of type $A_{N-1}$ possess $N=(2,0)$ supersymmetry. The former little string can also arise through the decoupling limit of a stack of N $NS5$  branes in type IIB string theory and the latter as the decoupling limit  of a stack of N NS5 branes in type IIA string theory. This dual description is a result of T-daulity that $S^1$-compactified little strings enjoy.\\
In 11-dimensional lift to M-theory one considers a configuration of M5-branes placed along a compactified transverse direction with M2-branes stretching between consecutive M5 branes. The BPS excitation of little strings are created by the stretched  open M2-branes. For N separated M5 branes along $S^1$ there are N intervals and that grades the BPS excitations with $U(1)^N$ quantum numbers. The compactness of the transverse direction makes it evident that little strings are the non-critical counterpart of closed fundamental string just like M-strings are the non-critical counterpart of open fundamental strings. Taking this analogy further the open M2 branes ending on M5-branes can combine to form  closed M2-branes and probe the bulk. This analogy breaks down in the decoupling limit where little string are confined to the M5-brane world volume and cannot move in the bulk of 11d space-time.\\
M-theory compactified on a torus $T^2$ is equivalent to Type IIB compactified on an $S^1$. In this picture the S-duality symmetry $SL(2,\mathbb{Z})$ of type IIB is interpreted as the modular group of $T^2$. The S-duality taken together with T-daulity combines into a non-perturbative symmetry of M-theory called U-dualty. This duality map can be used to give two equivalent gauge theory interpretations in type IIB strings of the M5-M2-branes configurations.
One gauge theory describes the Coulomb branch of a $U(N)$ gauge theory with massive bifundamental engineered by a configuration of one NS5-brane and N D5-branes. The second gauge theory is a circular quiver with N nodes of U(1) gauge theories  engineered by the S-dual configuration of one D5-brane and N NS5 branes.\\
On a further lift of the little strings description to F-theory, the type IIB D5-NS5 branes web is translated to a toric web diagram of a particular CY3-fold $X_N$ which is an elliptic fibration over the affine $A_{N-1}$ space, which itself is an elliptic fibration over the complex plane. The presence of two compactified directions in the toric web is translated as double fibration structure. The corresponding topological A-string partition function have been computed in \cite{Hohenegger:2015btj,Hohenegger:2016eqy} using topological vertex, M-string and Instanton calculus techniques. \\
We are interested in studying little strings on $K3\times T^2$. Given $N$ parallel M5-branes with world volume given by $K3\times T^2$ it is a standard result \cite{Minahan:1998vr} that the compactification of this theory on $T^2$ will result in $N=4$ $U(N)$ Super Yang Mills on $K3$. This relation indicates that the partition function of $N$ parallel M5-branes on $K3\times T^2$ is related to the partition function of $N=4$ Super Yang Mills on $K3$. In this work we want to
 compute the M5-branes partition function of $K3\times T^2$  which corresponds to the partition function of mass deformed  $N=2^*$ $SU(N)$ gauge theory on K3.  \\
The expression for the  partition function $Z^{U(1)}_{K3\times T^2}(\rho,\sigma,\nu)$ of $K3\times T^2$ corresponding to the $U(1)$  theory  was proposed in \cite{Dijkgraaf:1996xw,Dijkgraaf:1996it,Dijkgraaf:2002ac} as the  the generating function of elliptic genera of symmetric product $Sym^n(K3)$ of $K$3s. The manifold  $Sym^n(K3)$  describes the configuration of instanton moduli space of a $K3$ surface.
We know  from these works  a general result  that expresses the orbifold elliptic genera of  $Sym^n(M)$ in terms of that of $M$ with $M$ being  a  $4d$ (hyper)K\"ahler manifold,
\bea\label{eq:DVV}
Z(M\times T^2;p,q,y)=\sum_{N=0}^{\infty}p^N\chi(S^NM;q,y)=\prod_{n>0,m\ge 0,l}(1-p^nq^my^l)^{-c(mn,l)}
\eea 
where $p=e^{2\pi i \rho},q=e^{2\pi i \tau},y=e^{2\pi i z}$ and 
\bea\label{eq:chiM}
\chi(M;q,y)=\sum_{m\ge 0,l}c(m,n)q^my^l.
\eea

For $M=K3$ we have for the elliptic genus
\bea\label{eq:k3tn}
\chi(K3,\tau,z)=24\chi(TN_1,\tau,z)=\sum_{h\ge 0,m\in\mathbb{Z}}24 c(4kl-m^2)e^{2\pi i(h\tau+mz)}
\eea
where the Taub-NUT elliptic genus is defined by
\bea
\chi(TN_1,\tau,z)=\sum_{h\ge 0,m\in\mathbb{Z}} c(4kl-m^2)e^{2\pi i(h\tau+mz)}
\eea
Roughly speaking the equation (\ref{eq:k3tn}) indicates that  the $K3$ surface under consideration is composed of $24$ Taub-NUT spaces $TN_1$.
The generating function of $\chi(TN_1,\tau,z)$ can be expressed in terms of the weight 10 automorphic form $\Phi_{10}(\rho,\sigma,\nu)$ of $Sp(2,\mathbb{Z})$ as \cite{Dijkgraaf:2007sw}
\bea\label{eq:DDMMV}
Z(\rho,\sigma,m)&=&\sum_N e^{2\pi i N\sigma}\chi_{\rho\nu}(TN^N_1/S_N)\nonumber\\
&=&e^{-\pi i(\rho+\sigma+\nu)}\prod_{(k,l,m)>0}(1-e^{2\pi i(k\rho+l\sigma+m\nu)})^{-c(4kl-m^2)}\nonumber\\
&=&\frac{1}{\Phi_{10}(\rho,\sigma,\nu)^{\frac{1}{24}}}
\eea

 The genus expansion of the  free energy is given by 
\bea
\mathcal{F}=\ln Z=\frac{1}{\epsilon_1\epsilon_2}\mathcal{F}_0+\mathcal{F}_1+...
\eea
The exponential of the genus-one part $\mathcal{F}_1$ has  the following  relationship to $Z^{U(1)}_{K3\times T^2}(\rho,\sigma,\nu)$
\bea
e^{24\mathcal{F}_1}&=&\frac{1}{\Phi_{10}(\rho,\sigma,\nu)}=Z^{U(1)}_{K3\times T^2}(\rho,\sigma,\nu)\nonumber\\
&=&\sum_N e^{2\pi i N\sigma}\chi_{\rho\nu}(K3^N/S_N)\nonumber\\
\eea

The proof of (\ref{eq:DVV}) involves the results about orbifold conformal field theory and is a natural generalisation of  orbifold Euler characteristics $\chi_0$ and orbifold $\chi_y$ genus \cite{Dijkgraaf:1996xw},
\bea
Z_{K3\times T^2}^{G}=\sum_{k=0}^{\infty}Q_{\tau}^r\chi_{ell}(\mathcal{M}_{G}(K3,r);Q_\rho,m)
\eea
For a compact manifold $M$ the elliptic genus $\chi_{ell}(M;Q_\rho,m)$ is a holomorphic function on $\mathbb{H}\times\mathbb{C}$ and is a  jacobi form of weight zero and index $\frac{d}{2}$ when $c_1(M)=0$, where $\mathbb{H}$ denotes the complex upper half plane and $dim_{\mathbb{C}}(M)=d$. But if $c_1(M)$ is turned on the elliptic genus remains no more a modular form. However, as shown in \cite{Gritsenko:1999nm}, by introducing an automorphic correction to the elliptic genus it becomes a weak jacobi form of weight 0, index $\frac{d}{2}$. After taking into account $c_1(M)\ne 0$ the elliptic genus can be written as
\bea
\chi_{ell}(M;Q_\rho,m)=(\frac{\theta_1(\tau,m)}{\eta(\tau)^3})^d\int_M P(M,q,y)W(M,q)
\eea
where $ P(M,q,y)$ is given by
\bea
 P(M,q,y)=exp\big(-\sum_{n\ge 2}\frac{\wp^{(n-2)}(\tau,m)}{(2\pi i)^n n!}\sum_{i=1}^dx_i^n   \big)
\eea
and $W(M,q)$, also called the Witten factor, is given by
\bea
W(M,q)=exp\big( -\sum_{k\ge 2}\frac{B_{2k}E_{2k}(\tau)}{2k (2k)!} \sum_{i=1}^dx_i^{(2k)} \big)
\eea
for $x_i$ defined as the roots of the tangent bundle $TM$,$q=e^{2\pi i \tau},y=e^{2\pi i m}$,$\wp^{(\tau,m)}$ is the Weierstrass $\wp$-function and is a jacobi form of weight 2, index 0, and $\wp^{(n-2)}(\tau,m)$ is the $(n-2)$-th derivative of $\wp^{(\tau,m)}$.\\
If the manifold $M$ is non-compact \cite{Hohenegger:2013ala} and admits a $U(1)$ action with isolated fixed points $(m_1,,,m_N)$, then $\chi_{ell}(M;Q_\rho,m)$ can be calculated using equivariant fixed point theorem as
\bea
\chi_{ell}(M;Q_\rho,m)=\sum_{n=1}^N(\frac{i\theta_1(\tau,m)}{\eta(\tau)^3})^d(q_{1,n}...q_{d,a})^{-1}P(M,q,y,q_{a,b})W(M,q,q_{a,b})
\eea
where the weights at the fixed points of the $U(1)$ action are denoted here by $q_{a,b}$ for $a=1,...,d$ and $b=1,...,N$.\\
Upon taking the limit $Q_\rho\to 0$ the elliptic genus reduces to the $\chi_y$ genus
\bea
Z_{K3\times S^1}^{G}=\sum_{k=0}^{\infty}Q_{\tau}^r\chi_{y}(\mathcal{M}_{G}(K3,r),m)
\eea
A further limit $y\to 0$  results in the generating function of the Euler characteristic of  $\mathcal{M}_{G}$.
\bea
Z_{K3}^{G}=\sum_{k=0}^{\infty}Q_{\tau}^r\chi_{0}(\mathcal{M}_{G}(K3,r))
\eea
where  $\mathcal{M}_{G}(M,r)$ is the moduli space of rank $r$ and gauge group $G$-instantons on K3.


The building block of our computation is the (type IIb little strings) partition function $Z_{X_{1,1}}$ of the manifold $X_{1,1}$ \cite{Hohenegger:2015btj,Hohenegger:2016eqy}. It can also be interpreted as the partition function of  single wound M-strings configuration. 
The corresponding free energy $F=logZ_{X_{1,1}}$ can be expanded in the spacetime equivariant parameters $\epsilon_1,\epsilon_2$. In the unrefined limit $\epsilon_1=-\epsilon_2=\epsilon$  we can consider its universal ($\epsilon$  independent) part, which  turns out to be the jacobi form of weight zero and index 1, $\phi_{0,1}(\tau,z)$.
The Exponential-lift\footnote{multiplicative-lift,exponential-lift and Borcherds lift are all synonymous}\cite{Borcherds1995,Borcherds:1996uda,Dabholkar:2006xa} of $\phi_{0,1}$ gives the partition function of a single M5-brane on $K3\times T^2$, $Z_{K3\times T^2}^{U(1)}$.\\
Moving on to the Calabi-Yau 3-fold $X_{N,1}$ which is the orbifold of $X_{1,1}$ by $\mathbb{Z}_{N}\times\mathbb{Z}$. The orbifold action is lifted to the K3 instanton moduli space and only the $\mathbb{Z}_N\times\mathbb{Z}$ invariant configurations will contribute \cite{2019arXiv190701535B}. The description of instanton moduli space in terms of the $\mbox{Hilb}^n[K3]$ will imply projecting out the contribution of those subschemes which are not invariant under $\mathbb{Z}_N$. We can again extract the $\epsilon$ independent part of the free energy corresponding to singly wound M-strings to find  Jacobi forms of weight zero and index N,  and take its Exponential-lift to get the  partition function of $N$ M5-branes on $K3\times T^2$, $Z_{K3\times T^2}^{SU(N)}$.
The partition functions $Z_{K3\times T^2}^{SU(N)}$ turn out to be genus $2$ Siegel modular forms. Similar modular forms have been constructed \cite{Govindarajan:2010fu} using additive-lift in the context of counting $\frac{1}{4}$-BPS states which are $\mathbb{Z}_M$ twisted in the CHL $\mathbb{Z}_N$-orbifold.  The Seigel(para)modular forms that are constructed by Exponential-lift of certain  modular forms, also called the seed modular forms, have a natural interpretation as the generating function of symmetric product of certain CFTs.\\
We give plan of the paper now. After reviewing the effective 2d/4d description of the M5-branes partition functions of $K3\times T^2$ in section \ref{effective}, we compute in section \ref{SEED} the seed partition functions. In section \ref{LSPF},  we compute the little string partition function of $K3\times T^2$ using Borcherds lift of the seed partition function for $N=1,2,3$. In section \ref{GW} we discuss the corresponding Gromov-Witten potentials. In the last section \ref{RST} we discuss the relevance of Ray-Singer torsion to the (para)modular forms. Some useful identities and review material is given in the appendices.

\vspace{2cm}
  
\section{The effective 2d/4d \`{a} la  Vafa-Witten limit of the M5-branes Partition function }\label{effective}
Consider the setup of N parallel M5 branes wrapped on a six dimensional submanifold $M_4\times T^2$ of the 11d spacetime. If parallel M5 branes are compactified on $T^2$, it gives rise to effective $U(N) N=4$ supersymmetric Yang-Mills theory on $M_4$. It means that the partition function of $N$ M5 branes on $M_4\times T^2$, in the limit of small $T^2$ volume, is the same as the partition function of $U(N),N=4$ Super Yang Mills on $M_4$ \cite{Minahan:1998vr}. There is  another dual description in which N parallel M5 branes are wrapped on $M_4$ and in the small volume limit it will  give rise to  string degrees on freedom on $T^2$.
The gauge coupling constant is given by the complex structure $\tau$ of $T^2$.i.e $\tau=\frac{4\pi i}{g^2}+\frac{\theta}{2\pi}$. \\
A single M5-brane gives rise to the $U(1)$ theory. The world volume self-dual  field strength $B_{mn}$ gives rise to $19$ left moving  and 3 right moving  periodic bosonic fields. There are 3 non-compact bosonic fields both for left- and right-movers. Two more scalars both left and right movers transform as sections of  trivial canonical line bundle of  K3. In total we have 24 left moving and 8 right moving bosonic fields. The 
fermonic  field content comprise of 8 right-moving fields. In the heterotic dual frame this is the field content of heterotic string. In the presence of fermionic zero modes the worldsheet degrees of freedom can be quantified by the   regularised  elliptic genus
 \bea
 Z=\frac{\tau_2^{5/2}}{V_5}\mbox{Tr}[(-1)^{F_R}F_R^4q^{L_0}\bar{q}^{L_0}]
 \eea
 where the prefactor $\frac{\tau_2^{5/2}}{V_5}$ denotes the contribution of five uncompactified transverse bosons.\\
A $K3$ surface has $SU(2)$ holonomy and vanishing first Chern class. Due to the absence of an explicit expression for a metric on $K3$, orbifold models were extensively used to study certain types of topological invariants. 
Elliptic genus, for example, is independent of the complex structure moduli of the $K3$ surface and can equivalently be computed for the orbifold limit of  the $K3$ surface.  In the orbifold limit a $K3$ surface is modelled  by moding out $T^2\times T^2$ by the symmetry groups $\mathbb{Z}_1,\mathbb{Z}_2,\mathbb{Z}_3,\mathbb{Z}_4$, with orbifold singularities appearing at the fixed points of 
\bea
z_1\to e^{\frac{2\pi i}{n} }z_1,\quad z_2\to e^{-\frac{2\pi i}{n} }z_2
\eea
for each $n=1,2,3,4$. There exists orbifold invariant 2-forms $dz_1\wedge dz_2$,$d\bar{z}_1\wedge d\bar{z}_2$ and $dz_1\wedge d\bar{z}_1-dz_2\wedge d\bar{z}_2$ which are acted on by an $SU(2)$ symmetry. This allows the  $\mathcal{N}=4$ algebra to enter this picture .\\
On the $K3$ surface the $SU(2)$ instanton moduli space  is  $8.k-12$ dimensional \cite{Vafa:1994tf} for instanton number $k$. The description of instanton moduli space can be given in terms of the space of configuration of $2k-3$ distinct unordered points on $K3$. This configuration depends on $8k-12$ parameters and is hyper-K\"{a}hler. The corresponding moduli space has orbifold singularities and a resolution is necessary to give a compactified instanton moduli space. An equivalent description can be given in terms of  $\mbox{Sym}^{2k-3}(K3)$ which upon resolution of orbifold singularities is called  $\mbox{Hilb}^{2k-3}(K3)$, the Hilbert scheme of $2k-3$ points on $K3$. The analysis  for $\mbox{Sym}^{2k-3}(K3)$  and $\mbox{Hilb}^{2k-3}(K3)$ agree  because $K3$ is a hyperk\"{a}hler manifold. Unlike a general four manifold on which the holonomy is $SU(2)_L\times SU(2)_R$, $K3$ has $SU(2)_R$ holonomy and hence making a twist using $SU(2)_L$ does not differentiate between twisted and physical theories.  In the process of resolving the singularities the hyper-k\"{a}hler structure must be preserved and Betti numbers must also remain invariant.  The above description of  $SU(2)$ instanton moduli space is only valid for $k$ odd. However the partition function for general k can be predicted using modular constraints from S-duality.  For $SO(3)\simeq SU(2)/\mathbb{Z}_2$ there is no constraint on the allowed value of the instanton number $k$. We will restrict ourselves to the $SU(N)$. The partition function of K3 corresponding to a single M5-brane is given by \cite{Minahan:1998vr}
\bea
 Z_{U(1)}=G(q)=\frac{1}{\eta(\tau)^{24}}
\eea
with $q=e^{2\pi i\tau}$ and $\eta(\tau)$  being the Dedekind eta function. The $N=4$ $SU(2)$ gauge theory partition function of $K3$ corresponding to two M5-branes is given by
\bea\label{eq:zsu2}
Z_{SU(2)}&=&\frac{1}{4}G(q^2)+\frac{1}{2}G(q^{\frac{1}{2}})+\frac{1}{2}G(-q^{\frac{1}{2}})
\eea
This result has a nice interpretation in terms of M5-branes. In this interpretation $Z_{SU(2)}$ can alternatively be viewed as corresponding to single M5-brane wrapped two times over $K3\times T^2$. This can be done by considering the world volume of two M5-branes to be $K3\times T^2$ but now with $T^2$ double covering the original spacetime $T^2$. There are then three inequivalent choices of complex structures that we can have
\bea
\tilde{\tau}=2\tau,\frac{\tau}{2},\frac{\tau+1}{2}
\eea
The three terms in the partition function (\ref{eq:zsu2}) correspond to these three complex structures.\\
We can gneralize this picture to the case of $N$ parallel M5-branes wrapping $K3\times T^2$ with the corresponding $U(N)$ gauge theory on $K3$. In this case $T^2$ covers the original spacetime $T^2$ $N$ times and the inequivalent choices of complex structures are enumerated as follows: given a basis of 1-cycles on $T^2$, the following set of $GL(2)$ transformations enumerate the inequivalent choices of n-fold covers
\bea\label{eq:matabc}
\begin{pmatrix}
a & b \\
0 & d 
\end{pmatrix}
\eea
with $ad=N$,$b<d$ and $a,b,d\ge 0$.
 The $N=4$ $SU(N)$ gauge theory partition function of K3, $Z_{SU(N)}$, corresponding to $N$ parallel M5-branes can be written as an  Hecke transform  of $Z_{U(1)}$ in terms of the above matrices (\ref{eq:matabc}) as
\bea\label{eq:almostHecke}
Z_{SU(N)}=\frac{1}{N^2}\sum_{\substack{0\le a,b,d\in \mathbb{Z} \\ ad=N,b<d}}d G(\frac{a\tau+b}{d})
\eea
It is a  Hecke transformation of order $N$. 
\section{M2-M5 branes and the seed functions}\label{SEED}
We begin by  writing  down the relevant M-string partition function for M-theory vacuum  described by the $11d$ space-time $\mathbb{T}^2\times\mathbb{R}^3_{||}\times \mathbb{S}^1\times \mathbb{S}^1\times \mathbb{R}^4_{\bot}$. The following table gives the coordinate labels  and specifies the world volume directions of the BPS M5-M2-M-string configuration. \\

\vspace{2cm}
\begin{tabular}{ |p{2cm}|p{10cm}|  }
 \hline
 \multicolumn{2}{|c|}{11D M-theory space-time} \\
 \hline
 & $x^0\quad x^1 \quad x^2\quad x^3\quad x^4\quad x^5\quad x^6\quad x^7\quad x^8\quad x^9\quad x^{10}$\\
 \hline
 M5   & $\times\quad \times \quad \times\quad \times\quad \times\quad \times\quad \quad \quad \quad \quad $ \\
 M2&   $\times\quad \times \quad \quad \quad \quad \quad \quad \quad \qquad\times \quad $ \\
 M-string &$\times\quad \times \quad \quad \quad \quad \quad \quad \quad \quad \quad $ \\
 \hline
\end{tabular}
\vspace{2cm}

where $T^2$ spans the $\{x^1,x^2\}$ directions and $S^1\times S^1$ spans $x^5,x^6$ directions respectively, with rest of the directions being non-compact.
Below we summarise our notation which follows 
\cite{Hohenegger:2016eqy,Hohenegger:2015btj,Hohenegger:2015cba}
\bea
&x^1\sim x^1+2\pi R_1;\quad2\pi i R_1:=\tau; \quad Q_{\tau}=e^{2\pi i\tau},\nonumber\\
&x^6\sim x^6+2\pi R_6;\quad2\pi i R_6:=\rho;\quad  Q_{\rho}=e^{2\pi i\rho},\nonumber\\
&q=e^{2\pi i \epsilon_1}\quad  t=e^{-2\pi i \epsilon_2}
\eea
The M5-branes are placed at the following positions
\bea
0\le a_1\le a_2\le ...\le a_N\le 2\pi R_6
\eea
resulting in the corresponding gauge theory Coulomb branch parameters
\bea
t_{f_1}&=&a_2-a_1,\quad t_{f_2}=a_3-a_2,...,\quad t_{f_{N-1}}=a_N-a_{N-1},\nonumber\\t_{f_N}&=&2\pi R_6-\sum_{i=1}^{N-1}t_{f_i}=-i\rho-(a_N-a_1).
\eea
The N-tuple of integers $(k_1,k_2,...,k_N)$ specifies the distribution of $K=k_1+k_2+...+k_N$ $M2$-branes amongst the N intervals. To render the computation on the non-compact part of the manifold regularised one introduces the so-called $\Omega$-deformation as
\bea
\quad (z_1,z_2)&\to& (e^{2\pi i \epsilon_1}z_1,e^{2\pi i \epsilon_2}z_2)\nonumber\\
(\omega_1,\omega_2)&\to& (e^{2\pi i m-\pi i (\epsilon_1+\epsilon_2)}\omega_1,e^{-2\pi i m-\pi i (\epsilon_1+\epsilon_2)}\omega_2)\\
\eea
where $(\omega_1,\omega_2)=(x_7+i x_8,x_9+i x_{10})$ and $(z_1,z_2)=(x_2+i x_3,x_4+i x_{5})$.
From transverse space-time viewpoint  $m$ is the mass deformation parameter.
 The partition function can be computed using topological vertex formalism  \cite{Iqbal:2007ii}. Since it is simpler  to study the modular properties of the free energy,  below we give  its expression for the choice of vertical direction in the toric diagram as the preferred one. This will correspond to the type IIb little string partition function.
\begin{figure}
\begin{tikzpicture}[every node/.append style={midway}]
\draw[black, thick] (10,4) -- (10.6,4)node[anchor=west] {\quad $1$};
\draw[black, thick] (10,3) -- (9.4,3) node[anchor=east] {$2\quad$};
\draw[black, thick] (10,4) -- (10,3);
\draw[black, thick] (10,4) -- (9.5,4.5);
\draw[black, thick] (9.5,4.5) -- (9.5,5.1)node[anchor=south] {\quad $a$};
\draw[black, thick] (9.5,4.5) -- (8.9,4.5)node[anchor=east] {$1\quad$};
\draw[black, thick] (10,3) -- (10.5,2.5);
\draw[black, thick] (10.5,2.5) -- (11.1,2.5)node[anchor=west] {\quad $2$};
\draw[black, dotted] (10.5,2.5) -- (10.5,1.7);
\draw[black, dotted] (10.5,2.5) -- (10.5,1.7);
\draw[black, thick] (10.5,1.7) -- (11.1,1);
\draw[black, thick] (10.5,1.7)-- (9.8,1.7)node[anchor=east] {$N\quad$};
\draw[black, thick] (11.1,1)-- (11.7,1)node[anchor=west] {\quad $N$};
\draw[black, thick] (11.1,1)-- (11.1,0.3)node[anchor=north] {\quad $a$};
\draw[black,<->]      (6.9,0.4)   -- (6.9,5.1)node[right]{$\rho$};
 \draw[black,<->]       (9,0.1)   -- (11.8,0.1)node[below]{$\tau$};
\draw[black,<->]      (6.5,0.4)   -- (6.5,1.6)node[left]{$t_{f_N}$};
\draw[black,<->]      (6.5,2.9)   -- (6.5,4.4)node[left]{$t_{f_1}$};
\end{tikzpicture}
 \caption{Toric web diagram of the Calabi-Yau 3fold $X_{N,1}$. The web is compactified on a torus $S^1\times S^1$. $m$ is the K\"ahler parameter of the $\mathbb{P}^1$ which corresponds to the $(1,-1)$ line   }
    \label{N2}
\end{figure}
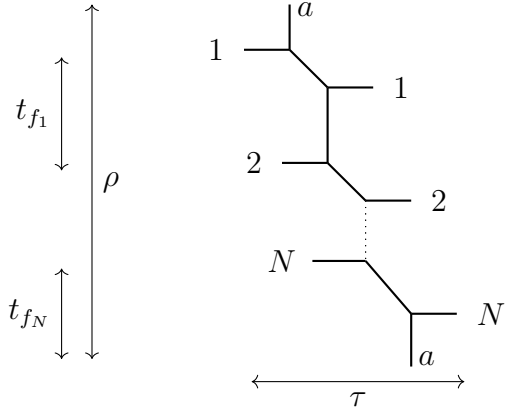
















For the vertical preferred direction, the topological string partition function 
$Z_{X_N}(\tau,m,t_{f_1},... ,t_{f_N},\epsilon_1,\epsilon_2)$ of the Calabi-Yau 3fold with toric web given in Fig.(\ref{N2})  can be expressed as  :
\bea
Z_{X_{N,1}} (\tau,m,t_{f_1},... ,t_{f_N} ,\epsilon_1,\epsilon_2) = Z_2(N,\tau,m,\epsilon_1,\epsilon_2)\tilde{Z}_N(\tau,m,t_{f_1},... ,t_{f_N} ,\epsilon_1,\epsilon_2)
\eea
where 

\bea
Z_2(N,\tau,m,\epsilon_1,\epsilon_2)=\mathcal{W}_{\emptyset\emptyset}(\tau,m,\epsilon_1,\epsilon_2)^N
\eea
and the open topological string wavefunction $\mathcal{W}_{\nu_i\nu_{i+1}}(Q_{\tau},Q_m,t,q)$ is defined by
\bea\label{opentopw}
&\mathcal{W}_{\nu_i\nu_{i+1}}(Q_{\tau},Q_m,t,q)=t^{-\frac{||\nu_{m+1}||^2}{2}}q^{-\frac{||\nu_{m}||^2}{2}}
\tilde{Z}_{\nu^t_m}(q^{-1},t^{-1})\tilde{Z}_{\nu_{m+1}}(t^{-1},q^{-1})Q_m^{-\frac{|\nu_m|+|\nu_{m+1}|}{2}}\prod_{k=1}(1-Q_{\tau}^k)^{-1}\nonumber\\&\times\prod_{i,j=1}\frac{(1-Q_{\tau}^kQ_m^{-1}q^{\nu_{m+1,i}-j+\frac{1}{2}}t^{\nu_{m,j}-i+\frac{1}{2}})(1-Q_{\tau}^{k-1}Q_m q^{\nu_{m,i}-j+\frac{1}{2}}t^{\nu_{m+1,j}^t-i+\frac{1}{2}})}{(1-Q_{\tau}^k q^{\nu_{m+1,i}-j+1}t^{\nu_{m+1,j}-i})(1-Q_{\tau}^{k} q^{\nu_{m,i}-j}t^{\nu_{m,j}^t-i+1})}\nonumber\\
\eea
where $\nu_i$ are partitions whose definitions are given in the appendix \ref{sec:Par}.\\
The non-perturbative part $\tilde{Z}_N(\tau,m,t_{f_1},... ,t_{f_N} ,\epsilon_1,\epsilon_2)$ of   the topological string partition function can be readily expressed by 
\bea\label{eq:w1}
\tilde{Z}_N(\tau,m,t_{f_1},... ,t_{f_N} ,\epsilon_1,\epsilon_2)=\sum_{\nu_1,\nu_2,...,\nu_N}Q_1^{|\nu_1|}Q_2^{|\nu_2|}...Q_N^{|\nu_N|}Z_{\nu_1\nu_2...\nu_N}(\tau,m,\epsilon_1,\epsilon_2)
\eea
where $Q_a=e^{-t_a}$ and $Z_{\nu_1\nu_2...\nu_N}(\tau,m,\epsilon_1,\epsilon_2)$ is defined as
\bea
Z_{\nu_1\nu_2...\nu_N}(\tau,m,\epsilon_1,\epsilon_2)=\prod_{a=1}^N\prod_{(i,j)\in\nu_a}\frac{\theta_1(\tau,z_{ij}^a)\theta_1(\tau,v_{ij}^a)}{\theta_1(\tau,w_{ij}^a)\theta_1(\tau,u_{ij}^a)}
\eea
and
\bea
z_{ij}^a&=&m-\epsilon_1(\nu_{a,i}-j+\frac{1}{2})+\epsilon_2(\nu_{a+1,j}^t-i+\frac{1}{2}),\nonumber\\
w_{ij}^a&=&-\epsilon_1(\nu_{a,i}-j+1)-\epsilon_2(\nu_{a+1,j}^t-i),\nonumber\\
 v_{ij}^a&=&-m-\epsilon_1(\nu_{a,i}-j+\frac{1}{2})+\epsilon_2(\nu_{a-1}^{t}-i+\frac{1}{2}),\nonumber\\ 
 u_{ij}^a&=&\epsilon_1(\nu_{a,i}-j)-\epsilon_2(\nu_{a,j}^t-i+1).
\eea
 A dual description can be given as the  five-dimensional $U(1)^N$ affine quiver gauge theory. The instanton moduli space corresponding to this gauge theory is given by $\mbox{Hilb}^{k_1}[\mathbb{C}^2]\times \mbox{Hilb}^{k_2}[\mathbb{C}^2]\times... \mbox{Hilb}^{k_N}[\mathbb{C}^2]$, where $\mbox{Hilb}^{n}[\mathbb{C}^2]$ is the Hilbert scheme of $n$ points on $\mathbb{C}^2$. 
The free energy admits the following expansion \cite{Hohenegger:2016eqy,Hohenegger:2015btj,Hohenegger:2015cba}
\bea
\Omega_N(\tau,m,t_{f_1},t_{f_1},...,t_{f_N},\epsilon_1,\epsilon_2)&=&\mbox{PLog}\tilde{Z}_{N}(\tau,m,t_{f_1},t_{f_1},...,t_{f_N},\epsilon_1,\epsilon_2)\nonumber\\&=& \sum_{k_1,,,,.k_{N-1}}Q^{k_1}_{f_1}...Q^{k_{N}}_{f_{N}}G^{k_1,...,k_{N}}(\tau,m,\epsilon_1,\epsilon_2)\nonumber\\
\eea
 The BPS configurations of the type IIb little string with charges $(k_1,k_2,...,k_N)$ are encoded by the functions\\ $G^{k_1,...,k_{N}}$$(\tau,m,\epsilon_1,\epsilon_2)$.
\subsection{Universal ($\epsilon_1,\epsilon_2$ independent) part  of the free energy  $G^{(1,1,...1)}(\tau,m,\epsilon_1,\epsilon_2)$}

For our purpose we are interested in the singly wound configuration of M-strings characterized by $G^{(1,1,...1)}(\tau,m,\epsilon_1,\epsilon_2)$  \cite{Hohenegger:2015btj}
\bea
G^{(1,1,...1)}(\tau,m,\epsilon_1,\epsilon_2)=W(\tau,m,\epsilon_1,\epsilon_2)^{N-1}\big[G^{(1)}(\tau,m,\epsilon_1,\epsilon_2)+(N-1)F^{(1)}(\tau,m,\epsilon_1,\epsilon_2)\big]
\eea
where 
\bea
W(\tau,m,\epsilon_1,\epsilon_2)&=&\frac{\theta_1(\tau;m+\epsilon_-)\theta_1(\tau;m-\epsilon_-)}{\theta_1(\tau;\epsilon_1)\theta_1(\tau;\epsilon_2)}-\frac{\theta_1(\tau;m+\epsilon_+)\theta_1(\tau;m-\epsilon_+)}{\theta_1(\tau;\epsilon_1)\theta_1(\tau;\epsilon_2)}\nonumber\\
G^{(1)}(\tau,m,\epsilon_1,-\epsilon_2)&=&F^{(1)}(\tau,m,\epsilon_1,\epsilon_2)=\frac{\theta_1(\tau;m+\epsilon_+)\theta_1(\tau;m-\epsilon_+)}{\theta_1(\tau;\epsilon_1)\theta_1(\tau;\epsilon_2)}
\eea
The importance of the function $G^{(1,1,...1)}$ stems from the fact that its universal part provides the seed function for the Exponential-lifting.
The partition functions of $K3\times T^2$ are obtained by the Exponential-lifting \cite{Borcherds1995,Borcherds:1996uda,Dabholkar:2006xa} of these seed functions. The seed function is obtained by taking the $\epsilon_1,\epsilon_2$ independent part of  $G^{(1,1,...1)}(\tau,m,\epsilon_1,\epsilon_2)$ and turns out to  be
\bea
\chi_N(\tau,m)&=&2^{-2N-1}3^{-N}(\varphi_{0,1}(\tau,m)+E_2(\tau)\varphi_{-2,1}(\tau,m))^{N-2}\nonumber\\ &&\bigg(-2(N-2)E_2(\tau)\varphi_{0,1}(\tau,m)\varphi_{-2,1}(\tau,m)-\nonumber\\
&&(N-1)(E_2(\tau)^2(N+2)-E_4(\tau)N)\varphi_{-2,1}(\tau,m)^2+2\varphi_{0,1}(\tau,m)^2\bigg)
\bigg)
\eea
It can be checked  that $\chi_N(\tau,m)$ is a  Jacobi form of weight $0$ and index $N$. To relate it to the computation of $K3\times T^2$ partition function we have to consider
its following multiple 
\bea\label{eq:multiple}
\psi_N(\tau,m)=24\chi_N(\tau,m)
\eea

The expression for first few Jacobi forms are as follows
\bea\label{eq:psi123}
\psi_1(\tau,m)&=&2\varphi_{0,1}(\tau,m)\nonumber\\
\psi_2(\tau,m)&=&\frac{1}{6}\varphi_{0,1}(\tau,m)^2-\frac{1}{3}E_2(\tau)^2\varphi_{-2,1}(\tau,m)^2+\frac{1}{6}E_4(\tau)\varphi_{-2,1}(\tau,m)^2\nonumber\\
\psi_3(\tau,m)&=&\frac{\varphi_{0,1}(\tau,m)^3}{72}-\frac{1}{12}E_2(\tau)^2\varphi_{0,1}(\tau,m)\varphi_{-2,1}(\tau,m)^2+\frac{1}{24}E_4(\tau)\varphi_{0,1}(\tau,m)\varphi_{-2,1}(\tau,m)^2\nonumber\\&+&\frac{5}{72}E_2(\tau)^3\varphi_{-2,1}(\tau,m)^3-\frac{1}{24}E_2(\tau)E_4(\tau)\varphi_{-2,1}(\tau,m)^3
\eea
The functions $\psi_{N}(\tau,m)$ are what we call the seed functions. The Borcherds-lift of these functions produce the $K3\times T^2$ partition functions.
This choice (\ref{eq:multiple}) is motivated by the fact that  the partition function of $U(1)$ theory is related to  the Borcherds-lift of $24\chi_1(\tau,m)=2\varphi_{0,1}(\tau,m)$\cite{Dijkgraaf:1996xw,Dijkgraaf:1996it,Dijkgraaf:2002ac}. So for $N=1$ this choice is correct.\\ Based on this intuition our main proposal is that the Borcherds-lift of $24\chi_N(\tau,m)$  is related to $N$ parallel M5-branes partition functions $Z_{K3\times T^2}^{SU(N)}$, with mass deformation, for other values of $N$. It is important to mention that the space of jacobi forms of wight $0$ and index $N$ is finite dimensional and moreover it forms a ring structure \cite{Gritsenko:1999nm}.\\
\section{Borcherds Lift and the Little strings partition function}\label{LSPF2}
The crucial relationship between (\ref{eq:DVV}) and (\ref{eq:chiM}) is provided by the Borcherds-lift, that is a way to obtain genus two modular forms by applying Hecke transformation on certain genus one jacobi forms.
More concretely, consider a Jacobi form $J_{w,l}(\tau,m)$ of weight $w$ and index $l$ with Fourier expansion given by
\bea
J_{w,l}(\tau,m)=\sum_{e,s}C_{J_{w,l}}(e,s)Q_{\tau}^eQ_m^s
\eea
where $Q_{\tau}=e^{2\pi i \tau},Q_m=e^{2\pi i m}$. The Borcherds lift of the jacobi form $J_{0,l}(\tau,m)$ of weight zero and index $l$, is defined as
\bea\label{eq:Borcherds}
B[J_{0,l}](\rho,\tau,m)=Q_{\tau}^aQ_m^bQ_{\rho}^c\prod_{(d,e,s)>0}(1-Q_{\rho}^{ld}Q_{\tau}^eQ_m^s)^{C_{J_{0,l}(de,s)}}
\eea
where $Q_{\rho}=e^{2\pi i \rho}$, $a=\sum_{s\in\mathbb{Z}}\frac{C_{J_{0,l}(0,s)}}{24}$,$b=\sum_{s>0}\frac{C_{J_{0,l}(0,s)s}}{2}$ $c=\sum_{s>0}\frac{C_{J_{0,l}(0,s)s^2}}{4}$. It is a genus two modular form of weight $\frac{C(0,0)}{2}$ where $C(0,0)$ is the constant term in the Fourier expansion of the corresponding jacobi form.
The Borcherds product is multiplicative by definition. Given two weakly holomorphic Jacobi forms $J_1$, $J_2$
we have
\bea
B[J_1+J_2]=B[J_1]B[J_2]
\eea
\\
If we expand the Borcherds product in $Q_{\rho}$, each coefficient of $Q_{\rho}$ is a jacobi form. Although the above can be defined for jacobi form of any weight, however, for jacobi forms of weight $0$ the above expression can be related to Hecke transforms. The above product can be related to Hecke transforms as we now describe. Let $J_{0,l}(\tau,m)$ be a jacobi form of weight zero and index $l$. Then the n-th Hecke transform of $J_{0,l}(\tau,m)$ is defined as 
\bea
(T_nJ_{0,l})(\tau,m)=n^{k-1}\sum_{ad=n,a>0}\sum_{b=0}^{d-1}J_{0,l}(\frac{a\tau+b}{d},am)
\eea
After doing some calculations as described in the appendix (\ref{LSPF}) we have
\bea
\sum_{n\ge 1}Q_{\rho}^{ln}(T_nJ_{0,l})(\tau,m)=-\sum_{(d,e,s)}C_{J_{0,l}}(de,s)log(1-Q_{\rho}^{ld}Q_{\tau}^eQ_m^s)
\eea
for $s\in\mathbb{Z},d>0,e\ge 0$.\\
The zero-th Hecke operator acting on the Jacobi form $J_{0,l}(\tau,m)$ is defined by
\bea
T_0(J_{0,l})(\tau,m)=\frac{C_{J_{0,l}(0,0)}}{2}\zeta(1)-\sum_{(0,e,s)>0}C_{J_{0,l}(0,s)}log(1-Q_{\tau}^2Q_m^s)
\eea
where $\zeta(1)=\sum_{h>0}\frac{1}{h}$. Combining the last two equations we have
\bea 
\sum_{n\ge 0}Q_{\rho}^{ln}(T_nJ_{0,l})(\tau,m)=\frac{C_{J_{0,l}(0,0)}}{2}\zeta(1)-\prod_{(d,e,s)>0}C_{J_{0,l}}(de,s)log(1-Q_{\rho}^{ld}Q_{\tau}^eQ_m^s)
\eea 
By taking the exponential of both sides we obtain
\bea\label{eq:expprod}
e^{-\sum_{n\ge 0}Q_{\rho}^{ln}(T_nJ_{0,l})(\tau,m)}=e^{-\frac{C_{J_{0,l}(0,0)}}{2}\zeta(1)}\prod_{(d,e,s)>0}(1-Q_{\rho}^{ld}Q_{\tau}^eQ_m^s)^{C_{J_{0,l}}(de,s)}
\eea
Next we consider the following expansion of the product term
\bea\label{eq:Zexp}
\prod_{(d,e,s)>0}(1-Q_{\rho}^{ld}Q_{\tau}^eQ_m^s)^{C_{J_{0,l}}(de,s)}=\Psi(\tau,m)(1+Q_{\rho}^lZ^{(1)}_{J_{0,l}}(\tau,m)+Q_{\rho}^{2l}Z^{(2)}_{J_{0,l}}(\tau,m)+Q_{\rho}^{3l}Z^{(3)}_{J_{0,l}}(\tau,m)+...)
\eea
where $Z_{J_{0,l}}^{(i)}$ are jacobi forms.
By substituting the eq.(\ref{eq:Zexp}) in (\ref{eq:expprod}) and comparing the coefficients of $Q_{\rho}$ expansion
we get  the functions $\Psi_{\psi}(\tau,m),Z_{J_{0,l}}^{(1)},Z_{J_{0,l}}^{(2)},Z_{J_{0,l}}^{(3)}$ in the following form
\bea
\Psi_{\psi}(\tau,m)&=&\prod_{(0,e,s)>0}(1-Q_{\tau}^eQ_m^s)^{C_{J_{0,s}}}\nonumber\\
Z_{J_{0,l}}^{(1)}(\tau,m)&=&-J_{0,l}(\tau,m)\nonumber\\
Z_{J_{0,l}}^{(2)}(\tau,m)&=&\frac{1}{2}(J_{0,l}(\tau,m))^2-\frac{1}{2}[J_{0,l}(2\tau,2m)+J_{0,l}(\frac{\tau}{2},m)+J_{0,l}(\frac{\tau+1}{2},m)]\nonumber\\
Z_{J_{0,l}}^{(3)}(\tau,m)&=&\frac{1}{6}\bigg( -T_1(J_{0,l}(\tau,m))^3+6T_2(J_{0,l}(\tau,m))T_1(J_{0,l}(\tau,m))+6T_3(J_{0,l}(\tau,m))\bigg)
\eea
Next in this procedure we construct $Z_{J_{0,l}}^{(i)}$ as an expansion in the mass parameter $m$ by evaluating the Hecke transformations. In the last step we resum the series (\ref{eq:Zexp}) in the variables $\rho$ and $\tau$ to get a mass expansion that is modular covaraint in $\rho$ and $\tau$.
We illustrate this procedure by constructing Borcherds-lift in three examples involving the jacobi forms $\psi_1(\tau,m),\psi_2(\tau,m)$ and $\psi_3(\tau,m)$.
\subsection{Borcherds lift of $\psi_1(\tau,m)$:  Little string partition function corresponding to  $U(1)$  theory }
So far we have been discussing the case of an arbitrary jacobi form of weight zero and index $l$. Let us now specialize to the case $l=1$ and consider jacobi form of weight $0$ and index $1$, $\psi_1(\tau,m)$. From the following Fourier expansion
\bea
\psi_1(\tau,m)=\sum_{e,s\in\mathbb{Z}}C_{\psi_1}(e,s)Q_{\tau}^eQ_m^{s}=2Q_m+2Q_m^{-1}+20+\mathcal{O}(Q_{\tau})
\eea
we notice that $C_{\psi_1}(0,1)=C_{\psi_1}(0,-1)=2$ and $C_{\psi_1}(0,0)=20$ with all other $C_{\psi_1}(0,s)$ vanishing.This leads to $a=b=c=1$. The function $\Psi(\tau,m)$ as defined in (\ref{eq:Zexp}) is given for $\psi_1(\tau,m)$ by
\bea 
\Psi_{\psi_1}(\tau,m)&=&\prod_{e>0,s\in\mathbb{Z}}(1-Q_{\tau}^eQ_m^s)^{C_{\psi_1}(0,s)}\prod_{s<0}(1-Q_m^s)^{C_{\psi_1}(0,s)}\nonumber\\&=&-Q_{\tau}^{-1}Q_m^{-1}\theta_1(\tau,m)^2\eta(\tau)^{18}\nonumber\\
&=&Q_{\tau}^{-1}Q_m^{-1}\varphi_{-2,1}(\tau,m)\eta(\tau)^{24}\nonumber\\
\eea
where we have used the definition of jacobi form $\varphi_{-2,1}(\tau,m)$ of weight $2$, index $1$ defined in (\ref{eq:phi21}). Since $\varphi_{-2,1}(\tau,m)$ has weight $-2$ and $\eta(\tau)$ has weight $\frac{1}{2}$, $\Psi_{\psi_1}(\tau,m)$ has weight $10$. Using the mass-expansion of $\varphi_{0,1}(\tau,m)$ calculated in (\ref{eq:phi01})  we obtain the following expansion of $\psi_1(\tau,m)$
\bea 
\psi_1(\tau,m)=24-2E_2(\tau)m^2+(\frac{1}{12}E_2(\tau)^2+\frac{1}{12}E_4(\tau))m^4+...
\eea
The first Hecke transformation $T_1$ of $\psi_1(\tau,m)$ is proportional to $\psi_1(\tau,m)$. The second and third Hecke transformations of $\psi_1(\tau,m)$ give us the jacobi form of weight $0$ and index $2,3$ respectively. These Hecke operators expressions are given by
\bea\label{eq:Heckeseries1}
T_2\psi_1(\tau,m)&=&36-6E_2(\tau)m^2+\frac{1}{2}(E_2(\tau)^2+2E_4(\tau))m^4+...\nonumber\\
T_3\psi_1(\tau,m)&=&32-8E_2(\tau)m^2+(E_2(\tau)^2+\frac{11}{3}E_4(\tau))m^4+...
\eea
By the definition of the Borcherds product
\bea
B[\psi_1](\rho,\tau,m)=Q_{\tau}Q_mQ_{\rho}\prod_{(d,e,s)>0}(1-Q_{\rho}^{2d}Q_{\tau}^eQ_m^{s})^{C_{\psi_1(de,s)}}
\eea
We have already seen in (\ref{eq:Zexp}) the Borcherds product can be written in the following way as well
\bea
B[\psi_1](\rho,\tau,m)=Q_{\tau}Q_mQ_{\rho}\Psi_{\psi_1}(\tau,m)\big(1+Q_{\rho}Z_{\psi_1}^{(1)}(\tau,m)+Q_{\rho}^2Z_{\psi_1}^{(2)}(\tau,m)+Q_{\rho}^3Z_{\psi_1}^{(3)}(\tau,m)+...\big)
\eea
where $Z_{\psi_1}^{(i)}(\tau,m)$ are jacobi forms of weight $0$ and index $i$. $Z_{\psi_1}^{(2)}(\tau,m)$ and $Z_{\psi_1}^{(3)}(\tau,m)$ involve second and third Hecke series of $\psi_1(\tau,m)$. We know the Hecke expansion given in (\ref{eq:Heckeseries1}) and so we have the following expansions of $Z_{\psi_1}^{(1)}(\tau,m),Z_{\psi_1}^{(2)}(\tau,m)$\\ and $Z_{\psi_1}^{(3)}(\tau,m))$
\bea
Z_{\psi_1}^{(1)}(\tau,m)&=&-\psi_1(\tau,m)\nonumber\\
Z_{\psi_1}^{(2)}(\tau,m)&=&252-42E_2(\tau)m^2+(\frac{7}{2}E_2(\tau)^2+E_4(\tau))m^4+...\nonumber\\
Z_{\psi_1}^{(3)}(\tau,m)&=&-1472+368E_2(\tau)m^2+(-46E_2(\tau)^2-\frac{2}{3}E_4(\tau))m^4+...
\eea
Using the above expressions of $Z_{\psi_1}^{(i)}(\tau,m)$ and the expansion of $\varphi_{-2,1}(\tau,m)$ given in (\ref{eq:phi21}) we give first few terms of $B[\psi_1](\rho,\tau,m)$
\bea\label{eq:Bpsi1}
B[\psi_1](\rho,\tau,m)&=&m^2\eta(\tau)^{24}\eta(\rho)^{24}+\frac{m^4}{12}E_2(\tau)E_2(\rho)\eta(\rho)^{24}\eta(\tau)^{24}+\nonumber\\&+&m^6\eta(\tau)^{24}\eta(\rho)^{24}\bigg(\frac{1}{24}E_2(\tau)^2E_4(\rho)-\frac{13}{24}E_2(\tau) ^2E_2(\rho)^2+\frac{1}{24}E_4(\tau)E_2(\rho)^2\}\bigg)\nonumber\\
&+&m^8 \eta(\tau)^{24}\eta(\rho)^{24}\bigg( \frac{1}{373248}E_6(\rho)E_2(\tau)^3+\frac{91}{746496}E_2(\rho)^3E_2(\tau)^3\nonumber\\&+&\frac{7}{622080}E_6(\rho)E_2(\tau)E_4(\tau) \bigg)
+...\nonumber\\
\eea
By observing this series expansion we can guess the following generic form for $B[\psi_1](\rho,\tau,m)$
\bea
B[\psi_1](\rho,\tau,m)=\sum_{k\ge 1}\eta(\tau)^{24}\eta(\rho)^{24}M_{2k-2}(\tau)M_{2k-2}(\rho)m^{2k}
\eea
where $M_{2k-2}(\tau)$ is a quasi modular form of weight $2k-2$
\bea
M_{2k-2}(\tau)=\sum_{a,b,b>0 \atop 2a+4b+6c=2k-2}\chi_{abc}E_2(\tau)^aE_4(\tau)^bE_6(\tau)^c
\eea
where $\chi_{abc}$ are numerical coefficients.

The $B[\psi_1](\rho,\tau,m)$ is a Siegel modular form of weight 10 and is nothing other than the Igusa cusp form $\Phi_{10}(\rho,\sigma,m)$. 
So the mass deformed single M5-brane partition function of $K3\times T^2$  is given by
\bea
Z^{U(1)}_{K3\times T^2}(\rho,\tau,m)=\frac{1}{B[\psi_1](\rho,\sigma,m)}
\eea

\subsection{Borcherds lift of $\psi_2(\tau,m)$: Little string partition function corresponding to $SU(2)$ theory}
In in section we will discuss the Borcherds lift for Jacobi form of weight $0$ having index $2$, in particular $\psi_2(\tau,m)$ which is defined in (\ref{eq:psi123}). By substituting $l=2$ in (\ref{eq:Borcherds}) we have the following infinite product for the Borcherd-lift of $\psi_2(\tau,m)$
\bea
B[\psi_2](\rho,\tau,m)=Q_{\tau}^aQ_m^bQ_{\rho}^c\prod_{(d,e,s)>0}(1-Q_{\rho}^{2d}Q_{\tau}^eQ_m^s)^{C_{\psi_2}(de,s)}
\eea
where $a=\sum_{s\in\mathbb{Z}}\frac{C_{\psi_2}(0,s)}{24}$,$b=\sum_{s>0}\frac{C_{\psi_2}(0,s)s}{2}$,$c=\sum_{s\in\mathbb{Z}}\frac{C_{\psi_2}(0,s)s^2}{4}$.\\
In order to find the values of $a,b$ and $c$ we need to know the Fourier coefficient $C_{\psi_2}(de,s)$. The $\psi_2(\tau,m)$  has the following Fourier expansion
\bea
\psi_2(\tau,m)=\sum_{e,s\in\mathbb{Z}}C_{\psi_2}(e,s)Q_{\tau}^eQ_m^s=4Q_m+4Q_m^{-1}+16+...
\eea
We see that $C_{\psi_2}(0,1)=C_{\psi_2}(0,-1)=4$ and $C_{\psi_2}(0,0)=16$ with all other $C_{\psi_2}(0,s)$ identically zero. Hence $a=1$, $b=c=2$. Again by using the same method of decomposing the product (\ref{eq:Zexp}) we obtain the following expansion of Borcherds-lift of $\psi_2(\tau,m)$
\bea\label{eq:Bpsi2Zexp}
B[\psi_2](\rho,\tau,m)=Q_{\tau}Q_m^2Q_{\rho}^2\Psi_{\psi_2}(\tau,m)\big(1+Q_{\rho}^2Z_{\psi_2}^{(1)}(\tau,m)+Q_{\rho}^4Z_{\psi_2}^{(2)}(\tau,m)+Q_{\rho}^6Z_{\psi_2}^{(3)}(\tau,m)+...\big)
\eea
Here $\Psi_{\psi_2}(\tau,m)$is given by
\bea
\Psi_{\psi_2}(\tau,m)&=&\prod_{(0,e,s)>0}(1-Q_{\tau}^eQ_m^s)^{C_{\psi_2}(0,s)}\nonumber\\
&=&\prod_{e>0}(1-Q_{\tau}^eQ_m)^{4}(1-Q_{\tau}^eQ_m^{-1})^{4}(1-Q_{\tau}^e)^{16}(1-Q_m^{-1})^{4}
\eea
The product form on the right hand side of the above equation can be simplified to
\bea
\Psi_{\psi_2}(\tau,m)&=&=Q_{\tau}^{-1}Q_m^{-2}\theta_1(\tau,m)^4\eta(\tau)^{12}\nonumber\\
&=&Q_{\tau}^{-1}Q_m^{-2}\varphi_{-2,1}(\tau,m)^2\eta(\tau)^{24}
\eea
The modular form $\Psi_{\psi_2}$ has weight $8$.\\
We know the mass-expansion of $\varphi_{-2,1}(\tau,m)$ and $\varphi_{0,1}(\tau,m)$ as given in (\ref{eq:phi21},\ref{eq:phi01}), therefore by using these expansions we have the mass-expansion of $\psi_2(\tau,m)$
\bea
\psi_2(\tau,m)=24-4m^2E_2(\tau)+\frac{1}{3}E_4(\tau)m^4+...
\eea
The functions $Z_{\psi_2}^{(i)}(\tau,m)$ are jacobi form of weight $0$ index $2i$. To find the expansion of these functions we have to know the Hecke transformations of $\psi_2(\tau,m)$. The first Hecke transformation $T_1$ of $\psi_2(\tau,m)$ is proportional to $\psi_2(\tau,m)$. The second and third Hecke expansions of $\psi_2(\tau,m)$
are
\bea
T_2\psi_2(\tau,m)&=&36-12E_2(\tau)m^2+3E_4(\tau)m^4+...\nonumber\\
T_3\psi_2(\tau,m)&=&32-16E_2(\tau)m^2+\frac{28}{3}E_4(\tau)m^4+...
\eea 
The function $Z_{\psi_2}^{(1)}(\tau,m)$ is equal to $-\psi_2(\tau,m)$ and using $T_2\psi_2(\tau,m)$, $T_3\psi_2(\tau,m)$ the functions $Z_{\psi_2}^{(2)}(\tau,m)$,$Z_{\psi_2}^{(3)}(\tau,m)$ are given as
\bea
Z_{\psi_2}^{2}(\tau,m)&=&252-84E_2(\tau)m^2+(8E_2(\tau)^2+5E_4(\tau))m^4+...\nonumber\\
Z_{\psi_2}^{3}(\tau,m)&=&-1472+736E_2(\tau)m^2+(-144E_2(\tau)^2-\frac{64}{3}E_4(\tau))m^4+...
\eea
Using $\ref{eq:Bpsi2Zexp}$ and the last equations the first few terms of Borcherds product of $\psi_2(\tau)$ are
\bea
B[\psi_2](\rho,\tau,m)&=&m^4\eta(\tau)^{24}\eta(2\rho)^{24}-m^6\frac{1}{6}\eta(\tau)^{24}\eta(2\rho)^{24}E_2(\tau)E_2(2\rho)\nonumber\\&+&m^8 \eta(\tau)^{24}\eta(2\rho)^{24}\bigg(-\frac{1}{720}E_4(\tau)E_4(2\rho)+\frac{1}{72}E_2(\tau)^2E_2(2\rho)^2 \bigg)+...\nonumber\\
\eea

This is a Siegel modular form of weight 8. 
It describes the partition function of two parallel M5-branes on $K3\times T^2$ with mass deformation
\bea
Z^{SU(2)}_{K3\times T^2}(\rho,\tau,m)=\frac{1}{B[\psi_2](\rho,\tau,m)}
\eea
It is related to the $N=2^*$ $SU(2)$ gauge theory partition function of the $K3$ surface.
\subsection{Borcherds lift of $\psi_3(\tau,m)$: Little string partition function corresponding to $SU(3)$ theory }
In this section we will discuss the Borcherds-lift of $\psi_3(\tau,m)$ as we defined in the previous two sections for $\psi_1(\tau,m)$ and $\psi_2(\tau,m)$. Consider  $\psi_3(\tau,m)$ given in (\ref{eq:psi123}) and its Fourier expansion 
given by
\bea
\psi_3(\tau,m)=\sum_{e,s\in\mathbb{Z}}C_{\psi_3}(e,s)Q_{\tau}^eQ_m^{s}=6Q_m+6Q_m^{-1}+12+...
\eea
This shows that $C_{\psi_3}(0,0)=12,C_{\psi_3}(0,-1)=C_{\psi_3}(0,1)=6$ with all other $C_{\psi_3}(0,s)$ identically zero. This implies that $a=1,b=c=3$.\\
Proceeding as in the previous sections for the computation of $B[\psi_1](\rho,\tau,m)$ and $B[\psi_2](\rho,\tau,m)$ we first have
\bea
B[\psi_3](\rho,\tau,m)=Q_{\tau}Q_m^3Q_{\rho}^3\Psi_{\psi_3}(\tau,m)\big(1+Q_{\rho}^3Z_{\psi_3}^{(1)}(\tau,m)+Q_{\rho}^6Z_{\psi_3}^{(2)}(\tau,m)+Q_{\rho}^9Z_{\psi_3}^{(3)}(\tau,m)+...\big)
\eea
The function $\Psi_{\psi_3}(\tau,m)$ turns out to be
\bea\label{eq:Psi3}
\Psi_{\psi_3}(\tau,m)&=&\prod_{(0,e,s)>0}(1-Q_{\tau}^eQ_m^s)^{C_{\psi_3}(0,s)}\nonumber\\
&=&Q_{\tau}^{-1}Q_m^{-3}\varphi_{-2,1}(\tau,m)^3\eta(\tau)^{24}
\eea
$\Psi_{\psi_3}(\tau,m)$ has weight 6.
The mass-expansion of $\psi_3(\tau,m)$ is given by
\bea
\psi_3(\tau,m)=24-6E_2(\tau)m^2+\frac{1}{4}(3E_4(\tau)-E_2(\tau)^2)m^4+...
\eea
To compute $Z_{\psi_3}^{(i)}(\tau,m)$ we first have to compute the Hecke expansions of $\psi_3(\tau,m)$
\bea\label{eq:Heckepsi3}
T_2(\psi_3(\tau,m))&=&36-18m^2E_2(\tau)+m^4(6E_4(\tau)-\frac{3}{2}E_2(\tau)^2)+...\nonumber\\
T_3(\psi_3(\tau,m))&=&32-24E_2(\tau)m^2+(17E_{\tau}-3E_2(\tau)^2)m^4+...
\eea
Using these Hecke transformations we can give the expressions for $Z_{\psi_3}^{(i)}$ as follows
\bea\label{eq:Zpsi3}
Z_{\psi_3}^{(1)}(\tau,m)&=&-\psi_3(\tau,m)\nonumber\\
Z_{\psi_3}^{(2)}(\tau,m)&=&252-126E_2(\tau)m^2+(\frac{27}{2}E_2(\tau)^2+12E_4(\tau))+...\nonumber\\
Z_{\psi_3}^{(3)}(\tau,m)&=&-1472+1104E_2(\tau)m^2+(-294E_2(\tau)^2-62E_4(\tau))m^4+...
\eea
Using the results in equations \ref{eq:Psi3},\ref{eq:Heckepsi3},\ref{eq:Zpsi3} we can write down the m-expansion of $B[\psi_3](\rho,\tau,m)$ as follows
\bea
B[\psi_3](\rho,\tau,m)&=&-m^6\eta(\tau)^{24}\eta(3\rho)^{24}+\frac{m^8}{4}\eta(\tau)^{24}\eta(3\rho)^{24}E_2(\tau)E_2(3\rho)+...\nonumber\\
&+&m^{10}\eta(\tau)^{24}\eta(3\rho)^{24}\bigg(-\frac{1}{1152}E_2(\tau)^2E_4(3\rho)-\frac{35}{1152}E_2(\tau)^2E_2(3\rho)^2\nonumber\\&-&\frac{1}{1152}E_4(\tau)E_2(3\rho)^2+\frac{17}{5760}E_4(3\rho)E_4(\tau) \bigg)+...\nonumber\\
\eea

This is a Siegel modular form of weight 6. It describes the partition function of three parallel M5-branes on $K3\times T^2$ with mass deformation 
\bea
Z^{SU(3)}_{K3\times T^2}(\rho,\tau,m)=\frac{1}{B[\psi_3](\rho,\tau,m)}
\eea
and is related to $N=2^*$ $SU(3)$ partition function of the $K3$ surface.\\
By following the same procedure we can workout $B[\psi_N](\rho,\tau,m)$ and hence $Z^{SU(N)}_{K3\times T^2}(\rho,\tau,m)$ for other values of N.\\

\section{Gromov-Witten potentials }\label{GW}
The gauge/geometry correspondence \cite{Gopakumar:1998ki} allows us to compute gauge theory prepotential from the  type IIA topological string amplitude.  Compactifying  type IIA superstrings on a CY$3$-fold one finds, among other terms, F-terms
\bea
\int d^4x F_g(t_i)R^2_+F_+^{2g-2},\quad g\ge1
\eea
where $R_+,F_+$ are self dual Riemann tensor and graviphoton field strength. The constant background value $F_+=\lambda_s$ is used as a parameter for asymptotic expansion and is called topological string coupling constant. $F_g$ is the A-model topological string amplitude of the $3-$fold. These  A-twisted topological string amplitudes are also interpreted as the generating functions of the genus $g$ Gromov-Witten invariants and appear as integrals on the moduli spaces of genus g Riemann surfaces. An observer on the world sheet will  interpret $F_g$ as roughly the number of maps form genus g Rieman surface, possibly with boundary, to the CY$3$-fold. The presence of world sheet boundary translates to the presence of Largrangian-branes on the target CY.\\

The genus g  topological A-model amplitudes are given by \cite{Kawai:2000px}
\bea
F_0&=&F_0^0+\mathcal{F}_0+... =\frac{c_{abc}t_at_bt_c}{6}+\sum_{\beta\in H_2(X,\mathbb{Z})}N_{\beta}^0e^{-\int_{\beta}\omega},\nonumber\\
F_1&=&F_1^0+\mathcal{F}_1+...=\frac{\sum_{a=1}^{h^{1,1}}t_a\int_X c_2(X)\wedge \omega_a}{24}+\sum_{\beta\in H_2(X,\mathbb{Z})}N_{\beta}^1e^{-\int_{\beta}\omega},\nonumber\\
F_{g\ge 2}&=&\mathcal{F}_g+...=(-1)^g(\int_{\mathcal{M}_g}\lambda_{g-1}^3)\frac{\chi(X)}{2}+\sum_{\beta\in H_2(X,\mathbb{Z})}N_{\beta}^ge^{-\int_{\beta}\omega}
\eea
where  $H_2(X,\mathbb{Z})$ is spanned by the classes $\{\omega_1,\omega_2,...,\omega_{h^{1,1}}\}$,$\mbox{D}_a$ are 4-cycles dual to $\omega_a$, $N_{\beta}^g$ are genus g Gromov-Witten invariants and $\lambda_{g-1}$ is the $(g-1)th$ Chern class of the Hodge bundle over the moduli space of genus g curves $\mathcal{M}_g$ and 
\bea
\int_{\mathcal{M}_g}\lambda_{g-1}^3=\frac{|B_{2g}||B_{2g-2}|}{(2g)(2g-2)(2g-2)!}.
\eea 
with $B_{2g}$ being  the Bernoulli numbers.\\

In the large volume limit of the base \cite{Kawai:2000px} of certain elliptically and the $K3$ fibered Calabi-Yau $3$-folds, the corresponding Gromov-Witten potential  at genus $g$ can be written as a lifting of a Jacobi form $\varphi_{2g-2,m}(\tau,z)$ of weight $2g-2$ and index $m$.
\bea\label{eq:FEg}
\mathcal{F}_g:=\sum_{l\ge 0}p^l\varphi_{2g-2,m}|_{V_l}(\tau,z)=\frac{c_g(0,0)}{2}\zeta(3-2g)+\sum_{(l,n,\gamma)>0}c_g(ln,\gamma)Li_{3-2g}(p^lq^n\zeta^{\gamma})
\eea
where $Li_{r}(p^lq^n\zeta^{\gamma})$ is the usual Polylogarithm for $r>0$ and a rational function for $r\le 0$. The triplet $(l,n,\gamma)$ is said to be positive if (i) $l>0$ (ii) $l=0,n>0$ (iii) $l=n=0,\gamma>0$.\\
Notice that in our case the jacobi forms $\psi_i(\tau,m)$ have zero weights and hence only the genus one contribution will be there. This additive-lift  generates genus two modular forms.
\subsection*{GW-potential for $N=1$}
\bea
\psi_1(\tau,m)=\sum_{e,s\in\mathbb{Z}}c_{1}(e,s)Q_{\tau}^eQ_m^{s}=2Q_m+2Q_m^{-1}+20+...
\eea

\bea
F_1=\sum_{l=0}^{\infty}Q_{\rho}^l\psi_{1}|_{T_{l}}(\tau,m)=\frac{c_1(0,0)}{2}\zeta(1)+\sum_{(l,n,\gamma)>0}c_1(ln,\gamma)Li_{1}(Q_{\rho}^lQ_{\tau}^nQ_{m}^{\gamma})
\eea
\subsection*{GW-potential for $N=2$}
\bea
\psi_2(\tau,m)=\sum_{e,s\in\mathbb{Z}}c_{2}(e,s)Q_{\tau}^eQ_m^s=4Q_m+4Q_m^{-1}+16+...
\eea

\bea
F_1=\sum_{l=0}^{\infty}Q_{\rho}^l\psi_{2}|_{T_{l}}(\tau,m)=\frac{c_2(0,0)}{2}\zeta(1)+\sum_{(l,n,\gamma)>0}c_2(ln,\gamma)Li_{1}(Q_{\rho}^lQ_{\tau}^nQ_{m}^{\gamma})
\eea
\subsection*{GW-potential for $N=3$}
\bea
\psi_3(\tau,m)=\sum_{e,s\in\mathbb{Z}}c_{3}(e,s)Q_{\tau}^eQ_m^{s}=6Q_m+6Q_m^{-1}+12+...
\eea

\bea
F_1=\sum_{l=0}^{\infty}Q_{\rho}^l\psi_{3}|_{T_{l}}(\tau,m)=\frac{c_3(0,0)}{2}\zeta(1)+\sum_{(l,n,\gamma)>0}c_3(ln,\gamma)Li_{1}(Q_{\rho}^lQ_{\tau}^nQ_{m}^{\gamma})
\eea


\section{Modular forms, and Ray-Singer torsion as the discriminant of the mirror curve}\label{RST}
The Ray-Singer torsion is the analytic analogue of the Reidemeister torsion. Intuitively the Reidemeister torsion of a smooth manifold counts the number of closed orbits of the vector fields which are associated to the smooth maps $f:X\to S^1$. The maps $f$ must not have any critical points. 
Consider the differential operator $\bar{\partial}_V$ coupled with a vector bundle \cite{Bershadsky:1993cx,Cecotti:1992vy} $V$ on a K\"ahler manifold $M$
\bea
\bar{\partial}_V:\quad \wedge^p\bar{T}^*\otimes V\to\wedge^{p+1}\bar{T}^*\otimes V
\eea
where $p=0,1,...,\mbox{dim}(M)-1$.
The existence of a suitable norm and a compatible connection on $V$ is assumed. 
Formally the holomorphic Ray-Singer torsion of $V$ is defined as the regularised product of the determinants over the spectrum of the Laplacian  $\Delta_V=\bar{\partial}_V\bar{\partial}^{\dagger}_V+\bar{\partial}^{\dagger}_V\bar{\partial}_V$ acting on $\wedge^p\bar{T}^*\otimes V$
\bea
I(V)=\prod(\mbox{det}^{\prime}\Delta^{(p)})^{(-1)^{p}p}
\eea
where the prime indicates that the zero modes are projected out.  
The Ray-Singer torsion is independent of the choice of metric that goes into the definition of the Laplacian $\Delta_V^p$. 
A useful, although formal, integral representation of $I(V)$ is given  by
\bea\label{eq:IV}
\mbox{log}(I(V))=\int_{\epsilon}^{\infty}\frac{ds}{s}\mbox{Tr}^{\prime}(-1)^pp e^{-s \Delta_V}
\eea
Interestingly in analogy with the genus one free energy the holomorphic Ray-Singer torsion satisfies the Quillen  anomaly equation given by
\bea\label{eq:anomalyIV}
\partial\bar{\partial}[log \mbox{I}(V)]=\partial\bar{\partial}\sum_p(-1)^p\mbox{ log det }(g^{(p)})+2\pi i\int_M \mbox{Td}(T)\mbox{Ch}(V)|_{(1,1)}
\eea
where $T$ denotes the tangent bundle of $M$, $\mbox{det}g^{(p)}$ is determinant of the matrix $g^{(p)}$ whose entries are the inner products in the space of  the kernel of $\bar{\partial}_V$. This anomaly arises in dealing with the determinant of the differential operators that depends on a parameter. The lefthand side and the righthand side of the equation (\ref{eq:anomalyIV}) denote a $(1,1)$-form on the complex structure moduli space.\\
In terms of the operators of the twisted $N=2$ conformal theory the relevant quantity is the  index defined by
\bea
\mathcal{I}=\mbox{Tr}[(-1)^FF_LF_Rq^{H_L}\bar{q}^{H_R}]
\eea
where $q=e^{2\pi i \tau},\bar{q}=e^{2\pi i \bar{\tau}}$ for the torus period $\tau$,$2\pi H_L=L_0-c/24,2\pi H_R=\bar{L}_0-c/24$. Moreover the right and left Fermi numbers operator $F_R,F_L$ are conserved. We can construct topological string genus one free energy (with $SL(2,\mathbb{Z})$ covariance) by taking an average   over the fundamental domain $\mathcal{F}$ of the upper half plane
\bea
F_1=-4\int_{\mathcal{F}}\frac{d^2\tau}{\tau_2}\mbox{Tr}[(-1)^FF_LF_Rq^{H_L}\bar{q}^{H_R}]
\eea
where $\tau_2=\mbox{Im}(\tau)$. If the $N=2$ theory is a sigma model the total hamiltonian is the Laplacian $\Delta=\bar{\partial}\bar{\partial}^{\dagger}+\bar{\partial}^{\dagger}\bar{\partial}$, and the harmonic $(p,q)$-forms describe the susy vacua. 
 The action of the operators $F_L$ and $F_R$ is translated to the holomorphic and anti-holomorphic degrees of the differential forms denoted by $p$ and $q$ respectively, here. 
Combining the last equation with (\ref{eq:IV}), we get $F_1$ in terms of the holomorphic Ray-Singer torsion as
\bea\label{F1RST}
F_1=\frac{1}{2}\sum_q(-1)^q q \mbox{log}\mbox{I}(\wedge^qT^*)
\eea
For $K3\times T^2$
\bea
F_1&=&\frac{1}{2}\sum_{q=0}^{3}\sum_{p=0}^{3}(-1)^{p+q} pq \mbox{log}(\mbox{det}^{\prime}\Delta^{(p,q)})\nonumber\\
&=&\mbox{log}\frac{\mbox{det}^{\prime}(\Delta^{0,0})^{\frac{9}{2}}\mbox{det}^{\prime}(\Delta^{1,1})^{\frac{1}{2}}}{\mbox{det}^{\prime}(\Delta^{1,0})^{3}}
\eea
The CY3-fold $K3\times T^2$ is a self mirror manifold \cite{Harvey:1996ts} and therefore one can use the complex structure moduli or K\"ahler moduli.
From general considerations of topological strings, $F_1$ satisfies a holomorphic anomaly equation \cite{Bershadsky:1993ta}. If we denote by $G_{i\bar{}j}$ the Zamolodchilov metric and by $C_i$ respectively $C_{\bar{j}}$ the action of operator $\phi_i$ respectively $\phi_{\bar{j}}$ on the Ramond ground states then the holomorphic anomaly equation satisfied by $F_1$ is given by\cite{Dijkgraaf:2002yn}
\bea\label{eq:topF1}
\partial_{\bar{j}}\partial_{i}F_1&=&\mbox{Tr}(-1)^FC_iC_{\bar{j}}-\frac{1}{12}G_{ij}\mbox{Tr}(-1)^F\nonumber\\
&=&\mbox{Tr}(-1)^FC_iC_{\bar{j}}-\frac{1}{12}G_{ij}\chi(M)
\eea
In fact  the Quillen anomaly equation (\ref{eq:anomalyIV}) is identical to the holomorphic anomaly equation  (\ref{eq:topF1}) satisfied by the genus-one topological string free energy. \\
 The topological strings partition function of $K3\times T^2$  is a genus one partition function  that is expressible in terms of Ray-Singer analytic torsion.
For our purposes the important fact is that the Ray-Singer analytic torsion is the discriminant of the mirror  curve. 
As shown in \cite{Hollowood:2003cv,1999math......4092Y}, the mirror curves for the non-compact CY3-fold $X_{m,1}$ are the theta divisors on the abelian varieties $A_{\tau}=\mathbb{C}^g/\Lambda_{\tau}$ and are defined as
\bea
\Theta_{\tau}:=\{z\in A_{\tau};\theta(z,\tau)=0 \},\quad \Theta_m:=\{(u,z,\tau)\in\mathbb{P}(V_m)\times\mathbb{A}:\sum u_a\theta_{a,0}(z,\tau)=0 \}
\eea
where $V_m=\mathbb{C}^{m^g}$ is the space of coordinates $u_a$ and we can define a  projection map $\pi=id_{\mathbb{P}(V_m)}\times p:\mathbb{P}(V_m)\times\mathbb{A}\to\mathbb{P}(V_m)\times\mathbb{H}_g$ and 
\bea
\theta(z,\tau):&=&\sum_{m\in\mathbb{Z}^g}\mbox{exp}\bigg(\pi i \mbox{m}^t\tau \mbox{m}+2\pi i \mbox{m}^t z  \bigg)\nonumber\\
\theta_{a,b}(z,\tau)&=&\sum_{n\in\mathbb{Z}_g}\mbox{exp}\big(\pi i(n+a)^t\tau(n+a)+2\pi i(n+a)^t(n+b) \big)
\eea
for $a,b\in\mathbb{R}^g$ and $\mathbb{H}_g$ is the genus g Siegel upper half plane. For a review see the appendix \ref{sec:DRS}.
 The fibers $\Theta_{m,(u,\tau)}=\pi^{-1}(u,\tau)$ are projective spaces and are elements of a complete linear system \cite{Griffiths:433962} denoted by $|L_m,\tau|$. The singular locus or the discrimant locus of $\pi:\Theta_m\to\mathbb{P}(V_m)\times\mathbb{H}_g$ is defined by $\mathcal{D}_{g,m}:=\{(u,\tau)\in\mathbb{P}(V_m)\times\mathbb{H}_g:\mbox{Sing}\Theta_{m,(u,\tau)}\ne\emptyset \}$.
One of the main results proven in \cite{1999math......4092Y}, is that the Ray-Singer  torsion $I(\Theta_{\tau})$ is related to the Siegel cusp form of weight $\frac{(g+3)g!}{2}$  as
\bea
I(\Theta_{\tau})=\bigg((\mbox{detIm}\tau)^{\frac{(g+3)g!}{2}}|\Delta_g(\tau)|^2\bigg)^{\frac{(-1)^{g+1}}{(g+1)!}}
\eea
 Moreover $\Delta_g$  can be factorized as
\bea
\Delta_g(\tau)=\chi_g(\tau)J_g(\tau)^2
\eea
where $J_g(\tau)$ is a Siegel modular form of weight $\frac{(g+3)g!}{4}-2^{g-3}(2^g+1)$ and $\chi_g(\tau)$ corresponding to  $\theta_{null,g}$ can be written in terms of the characteristic theta constants as
\bea
\chi_g(\tau):=\prod_{(a,b) even}\theta_{a,b}(0,\tau)
\eea
A genus two theta constant is defined as
\bea
\theta_{a_1,a_2,b_1,b_2}\bigg(0,\begin{pmatrix}
\tau & 0 \\
0 & \tau 
\end{pmatrix}\bigg)=\theta_{a_1b_1}(0,\tau)\theta_{a_2b_2}(0,\tau)
\eea
There are ten even  theta constants  given by
\bea
\theta_{0000},\theta_{1000},\theta_{0100},\theta_{0010},\theta_{0001},\theta_{1100},\theta_{0110},\theta_{0011},\theta_{1001},\theta_{1111}
\eea
In many cases the modular forms can be expressed in terms of the theta constants, 
such as the Siegel theta constant $\Delta_{\frac{1}{2}}(\Omega)$ can be expressed as
\bea
\Delta_{\frac{1}{2}}(\Omega)^2=(\frac{1}{2}\theta_{1111}(\Omega^{\prime}))^2
\eea
for $\Omega=\begin{pmatrix}\tau & z\\ z&\sigma \end{pmatrix} $ and $\Omega^{\prime}=\begin{pmatrix}\tau &2 z\\ 2z&2\sigma \end{pmatrix} $. 

\section*{Acknowledgements}
The authors would like to thank Amer Iqbal for discussions and the Abdus Salam School of mathematical sciences, Lahore, Pakistan, where part of  the project was completed.


\appendix
\section{Modular forms and paramodular forms}
\subsection{Modular forms}
To begin with the simplest examples of modular forms \cite{eichler2013theory,Dabholkar:2012nd},  the Eisenstein series $E_k(\tau)$ for $k>2$  are modular forms of weight $k$
\bea
E_k(\frac{a\tau+b}{c\tau+d})=(c\tau+d)^kE_k(\tau)
\eea
under the $SL(2,\mathbb{Z})$ transformations  $\begin{pmatrix}
a & b \\
c & d 
\end{pmatrix}\in SL(2,\mathbb{Z})$  and for $\tau$ in the upper half plane $\mathcal{H}$.
The second Eisenstein series on the other hand transforms anomalously
\bea
E_2(\frac{a\tau+b}{c\tau+d})=(c\tau+d)^2E_2(\tau)-\frac{c(c\tau+d)}{4\pi i}
\eea
The modular forms $E_k,k>2$ form a ring which is generated by $\{E_4(\tau),E_6(\tau)\}$. \\
The Jacobi form $J_{w,l}$ of weight $w$ and index $l$ is a holomorphic function with the following properties
\bea
J(\frac{a\tau+b}{c\tau+d},\frac{m}{c\tau+d})&=&(c\tau+d)^{w}e^{\frac{2\pi i l c m^2}{c\tau+d}}J(\tau,m)\nonumber\\
J(\tau,m+u\tau+v)&=&e^{-2\pi il(u^2\tau+2um)}J(\tau,m)\nonumber\\
\eea
for $\begin{pmatrix}
a & b \\
c & d 
\end{pmatrix}\in SL(2,\mathbb{Z})$ and $u,v,l\in\mathbb{Z}$.\\
 Two important Jacobi forms of index 1 that are often used in the text are $\varphi_{-2,1}(\tau,m),\varphi_{0,1}(\tau,m)$. They can be expressed as
 \bea
 \varphi_{0,1}(\tau,m)&=&4\sum_{a=2}^4\frac{\theta_a(\tau,m)^2}{\theta_a(\tau,0)^2}=\frac{3}{\pi}\wp(\tau,m)\frac{\theta_1(\tau,m)^2}{\eta(\tau)^6}\nonumber\\
 \varphi_{-2,1}(\tau,m)&=&-\frac{\theta_1(\tau,m)^2}{\eta(\tau)^6}=-m^2exp\big(\sum_{k\ge 1}\frac{(-1)^kB_{2k}E_{2k}(\tau)}{(2k)!}m^{2k} \big)
 \eea
 where the Weirstrass $\wp(z,w)$ function is defined as
 \bea
 \wp(z,w)=\frac{1}{z^2}+\sum_{w\in\mathbb{Z}+\tau\mathbb{Z}\ne0}\big(\frac{1}{(z-w)^2}-\frac{1}{w^2} \big)
 \eea
 and the Bernoulli numbers $B_{2k}$ are defined by
 \bea
 \frac{x}{e^x-1}=\sum_{k\ge0}B_k\frac{m^k}{k!}
 \eea
 The Bernoulli number $B_k$ for k odd is zero.\\
 Theeta function $\eta(\tau)$ and Jacobi theta functions $\theta_i,i=1,4$ are given by
 \bea
 \eta(\tau)&=&e^{\frac{\pi i \tau}{12}}\prod_{n\ge 1}(1-Q_{\tau}^n)\nonumber\\
 \theta_1(\tau,m)&=&-iQ_{\tau}(Q_m^{-\frac{1}{2}}-Q_m^{\frac{1}{2}})\prod_{r\ge 1}(1-Q_{\tau}^r)(1-Q_{\tau}^rQ_m)(1-Q_{\tau}^rQ_m^{-1})\nonumber\\
 \theta_2(\tau,m)&=& Q_{\tau}^{\frac{1}{8}}\prod_{r\ge 1}(1-Q_{\tau}^r)(1+Q_{\tau}^rQ_m)(1+Q_{\tau}^rQ_m^{-1})\nonumber\\
  \theta_3(\tau,m)&=& \prod_{r\ge 1}(1-Q_{\tau}^r)(1+Q_{\tau}^{r-\frac{1}{2}}Q_m)(1+Q_{\tau}^{r-\frac{1}{2}}Q_m^{-1})\nonumber\\
 \theta_4(\tau,m)&=& \prod_{r\ge 1}(1-Q_{\tau}^r)(1-Q_{\tau}^{r-\frac{1}{2}}Q_m)(1-Q_{\tau}^{r-\frac{1}{2}}Q_m^{-1})  
 \eea
 for $Q_{\tau}=e^{2\pi i \tau}$, $Q_m=e^{2\pi i m}$. \\ The Jacobi theta function $\theta_1(\tau,m)$ satisfies the following important properties
 \bea
 \theta_1(\tau,m+\frac{\zeta}{2\pi})&=&\theta_1(\tau,m)exp\big( -\frac{1}{24}E_2(\tau)\zeta^2+\frac{\theta_1^{\prime}\zeta}{2\pi\theta_1(\tau,m)}-\sum_{n\ge 2}\wp^{(n-2)}(\tau,m)\frac{\zeta^n}{n!}  \big)\nonumber\\
 \theta_1(\tau,\frac{\zeta}{2\pi})&=&\zeta \eta^3(\tau)exp\big(\sum_{k\ge 1}(-1)^k\frac{C_{2k}}{2k(2k)!}E_{2k}(\tau)\zeta^{2k} \big)
 \eea
 $\varphi_{-2,1}(\tau,m)$ and $\varphi_{0,1}(\tau,m)$ are the two most important examples of jacobi forms of index $1$ and they are defined as
 \bea\label{eq:phi21}
 \varphi_{-2,1}(\tau,m)&=&-\frac{\theta_1(\tau,m)}{\eta(\tau)^6}\nonumber\\
 &=&-m^2exp\bigg(\sum_{k\ge 1}\frac{(-1)^kB_{2k}E_{2k}(\tau)}{k(2k)!} \bigg)\nonumber\\
 &=&-m^2+\frac{E_2(\tau)}{12}m^4
+\frac{-5E_2(\tau)^2+E_4(\tau)}{1440}m^6+...
\eea
\bea\label{eq:phi01}
\varphi_{0,1}(\tau,m)&=&4\sum_{i=2}^4
\frac{\theta_i(\tau,m)^2}{\theta_i(\tau,0)^2}\nonumber\\
&=&12-E_2(\tau)m^2+\frac{1}{24}(E_2(\tau)^2+E_4(\tau))m^4+...
\eea
  \subsection{Paramodular forms}
 The paramodular group of level t \cite{Belin:2016knb,Belin:2018oza,Belin:2019jqz} , $\Gamma_t$, is defined as
 \bea
 \Gamma_t=\begin{bmatrix}
\mathbb{Z} & t\mathbb{Z} & \mathbb{Z}&\mathbb{Z}\\
\mathbb{Z} & \mathbb{Z} & \mathbb{Z}&t^{-1}\mathbb{Z}\\
\mathbb{Z} & t\mathbb{Z} & \mathbb{Z}&\mathbb{Z}\\
t\mathbb{Z} & t\mathbb{Z} & t\mathbb{Z}&\mathbb{Z}
\end{bmatrix}\cap Sp(4,\mathbb{Q})
 \eea
 An extension of $\Gamma_t$ can be defined as
 \bea
 \Gamma_t^+=\Gamma_t\cup\Gamma_tV_t
 \eea
 where
 \bea
 V_t=\begin{bmatrix}
0 &t& 0&0\\
1 &0& 0&0\\
0 &0& 0&1\\
0 &0& t&0
\end{bmatrix}
 \eea
 A matrix $M$ of $\Gamma_t^+$ can be decomposed into block form as
 \bea
 M=\begin{pmatrix}
A &B\\
C &D
\end{pmatrix}
 \eea
 Recall that the Siegel upper half plane $\mathbb{H}_2$ is defined in terms of the matrix $\Omega=\begin{pmatrix}
\tau &z\\
z &\rho
\end{pmatrix}$
as 
\bea
\mbox{det}(\mbox{Im}(\Omega))>0,\quad \mbox{Tr}(\mbox{Im}(\Omega))>0
\eea
The action of $M$ on $\Omega$ is given by
\bea
M(\Omega)=(A\Omega+B)(C\Omega+D)^{-1}
\eea
A weight $k$ meromorphic paramodular form $\Phi_k(\Omega)$ satisfies
\bea
\Phi_k(M(\Omega))=\mbox{det}(C\Omega+D)^k\Phi_k(\Omega)
\eea
 \section{Partitions}\label{sec:Par}
 \label{Appendix B}
  A sequence $\nu=(\nu_1,...,\nu_k)$ with non-increasing order is called the partition of some non-negative integer. The length of the partition $\nu$ is denoted by $l(\nu)$ and size of the partition is denoted by $|\nu|$. A Young diagram  gives a pictorial representation of the partition $\nu$ as
  \bea
  \{(i,j)\in \mathbb{N}|1\le j\le \nu_i \}
  \eea
  The transpose of partition $\nu^t$ is obtained by by reflecting $\nu$ around the diagonal.
The arm length $a(\nu)$, leg length $l(\nu)$ and the hook length $h(\nu)$ are defined as
\bea
a(\nu)&=&\nu_i-j,\nonumber\\
l(\nu)&=&\nu_j^t-i,\nonumber\\
h(\nu)&=&\nu_i-j+\nu_j^t-i+1
\eea 
\section{Borcherds Lift}\label{LSPF}
For weight zero Jacobi forms the Borcherds product can be expressed in terms of the exponential lift of Hecke operator.
 To elaborate on this procedure \cite{Aoki2005SIMPLEGR,Dabholkar:2006xa,Eguchi:2011aj,Borcherds1995,Borcherds:1996uda} first consider the group $G_0(N)$ consisting of matrices with integer entries of the block-diagonal form
 \bea
\{ \  \left[ {\begin{array}{cc}
  A& B \\
  NC &D \\
  \end{array} } \right]\in Sp(2,\mathbb{Z})\}\supset \Gamma_0(N)
 \eea
 
  $\Gamma_0(N)$ admits an action of the Hecke operator $T_n$ for integer n, such that for a weak Jacobi form $\phi_{k,m}$ of weight $k$ and index $m$ , $T_n(\phi_{k,m})=\phi_{k,mn}$ is another weak Jacobi form of weight $k$ and index $mn$. Since $\Gamma_0(N)$ has multiple cusps in its fundamental domain the Hecke operator is   more involved as compared to that of $SL(2,\mathbb{Z})$.
For a given weak Jacobi form $\phi^k(\rho,z)$ we define 

\bea\label{eq:Mlift}
L\phi^k(\rho,\sigma,z)=\sum_{n=1}T_n(\phi^k)(\rho,z)e^{2\pi i \sigma n}
\eea

For  $L\phi^k(\rho,\sigma,z)$ to be a Siegel modular form of zero weight it must be invariant under modular transformation, elliptic transformation and must have a Fourier expansion. The modular form is defined on the Siegel upper half plane $\mathcal{H}_2$ defined as
$\mathcal{H}_2=\{\Omega\in\mathcal{M}_2(\mathbb{C})|\Omega=\Omega^T, Im(\Omega)>0  \}$.
Clearly  $T_n(\phi^k)(\rho,z)e^{2\pi i \sigma n}$ is invariant under the modular and elliptic transformations. To investigate the third condition consider writing $L\phi^k$ in a product form. Using the explicit form of Hecke transformation the equation (\ref{eq:Mlift}) can be written as
\bea
L\phi=\sum_{n=1}^{\infty}  \bigg(\frac{1}{n}\sum_{ \  \left[ {\begin{array}{cc}
  a& b \\
  c &d \\
  \end{array} } \right]\in \Gamma_0(N)/\Delta_N(n)} e^{-\frac{2\pi i m c z^2}{c\rho+d}}\phi(\frac{a\rho+b}{d},az)   \bigg)e^{2\pi i \sigma n}
\eea
To write a representative element of  $ \Gamma_0(N)/\Delta_N(n)$ choose the complete set of cusps $\{s\}$ of $\Gamma_0(N)$ given by a set of matrices $g_s$,
\bea
g_s\in SL(2,\mathbb{Z})= \  \left[ {\begin{array}{cc}
  x_s& y_s \\
  z_s &w_s \\
  \end{array} } \right]
\eea
Next define a natural number $h_s$ by
\bea
g_s^{-1}\Gamma_0(N)g_s\cap P(\mathbb{Z})=\{\pm \left[ {\begin{array}{cc}
  1& h_sn \\
  0 &1 \\
  \end{array} } \right];n\in\mathbb{Z} \}
\eea
where $P(\mathbb{Z})$ is the set of all upper-triangular matrices over the field of integers with unit determinant. A more explicit form for $ \Gamma_0(N)/\Delta_N(n)$ is then given by
\bea
 \Gamma_0(N)/\Delta_N(n)=\cup_s\{g_s=\left[ {\begin{array}{cc}
  a& b \\
  0 &d \\
  \end{array} } \right];a,b,d\in\mathbb{Z},\quad ad=n,\quad az_s=0\quad mod\quad N \nonumber\\, b=0,1,...,h_sd-1\}
\eea
For each cusp define $n_s=\frac{N}{g.c.d(z_s,N)}$ and the form
\bea
\phi_s(\rho,z)=\phi(\frac{x_s\rho+y_s}{z_s\rho+w_s},\frac{z}{z_s\rho+w_s})
\eea
that can be Fourier expanded as
\bea
\phi_s(\rho,z)=\sum_{n,l}c_s(n,l)e^{2\pi i (n\rho+lz)}
\eea
$c_s(n,l)=c_{s,l}(4n-l^2)$. Next following \cite{Dabholkar:2006xa}
\bea
L\phi&=&\sum_s\sum_{n=1}\sum_{ad=n,az_s=0 \quad mod\quad N}\sum_{b=0}^{h_sd-1}\phi_s(\frac{a\rho+b}{d},az)e^{2\pi i n\sigma}\nonumber\\
&=&\sum_s\sum_{n=1}\sum_{ad=n,a\in n_s\mathbb{Z}}(ad)^{-1}dh_s\sum_{n,l\in\mathbb{Z}}c_{s,l}(4n_1d-l^2)e^{2\pi i(an_1\rho+a l z+n\sigma)}\nonumber\\
&=&\sum_s h_s\sum_{a=1}\frac{1}{an_s}\sum_{m=1}\sum_{n,l\in\mathbb{Z}}c_{s,l}(4n_1d-l^2)e^{2\pi i(a n_1\rho+a lz+m\sigma)}\nonumber\\
&=&\sum_s\frac{h_s}{n_s}log\bigg(\prod_{l,m,n\in\mathbb{Z}|m\ge 1}(1-e^{n_s(n_1\rho+lz+m\sigma)})^{c_{s,l}(4mn_1-l^2)} \bigg)
\eea
or
\bea\label{eq:product}
e^{L\phi}=\prod_{(n,l,m)>0}(1-(q^ny^lp^m)^{n_s})^{h_sn_s^{-1}c_{s,l}(4mn-l^2)}
\eea
for $q=e^{2\pi i \rho},y=e^{2\pi i z},p=e^{2\pi i \sigma}$. In the product representation (\ref{eq:product}) although the coefficients $c_{s,l}(4mn-l^2)$ are symmetric under $m\leftrightarrow n$, the ranges of $m$ and $n$ are not. To make the expression symmetric we have to multiply it by an appropriate factor that can be determined by inspection. The result of this is that one obtains a Siegel modular form as the multiplicative-lift of the weak Jacobi form $\phi(\rho,z)$,
\bea\label{eq:Bordef}
\Phi_k=Q_{\tau}^aQ_m^bQ_{\rho}^c\prod_{(n,l,m)>0}(1-(Q_{\tau}^nQ_m^lQ_{\rho}^m)^{n_s})^{h_sn_s^{-1}c_{s,l}(4mn-l^2)}
\eea
for $b$ some integer and $a\ge0 ,c\ge 0$.   The condition $(l,m,n)>0$ means one of the following\\
 $(i) m>0$,$n,l\in\mathbb{Z}$ or \\
 $(ii)m=0,n>0,l\in\mathbb{Z}$\\
 $(iii)n=m=0,l<0$\\
and  $a=\sum_{s\in\mathbb{Z}}\frac{C_{J_{0,l}(0,s)}}{24}$,$b=\sum_{s>0}\frac{C_{J_{0,l}(0,s)s}}{2}$ $c=\sum_{s>0}\frac{C_{J_{0,l}(0,s)s^2}}{4}$. We can write this product as \cite{Gritsenko:1996tm}
\bea
Q_{\tau}^aQ_m^bQ_{\rho}^c\prod_{n>0,l\in\mathbb{Z}}(1-Q_{\tau}^nQ_m^l)^{c(0,l)}\prod_{m>0,n,l\in\mathbb{Z}}(1-Q_{\tau}^nQ_m^lQ_{\rho}^t)^{c(mn,l)}\prod_{n=m=0,l<0}(1-Q_m^l)^{c(0,l)}\nonumber\\
\eea
Next note that
\bea
Q_m^b\prod_{n=m=0,l<0}(1-Q_m^l)^{c(0,l)}=\prod_{l<0}Q_m^{-\frac{lc(0,l)}{2}}(1-Q_m^{l})^{c(0,l)}=\prod_{l>0}(Q_m^{\frac{l}{2}}-Q_m^{\frac{-l}{2}})^{c(0,l)}
\eea
Then
\bea
&&Q_{\tau}^aQ_{m}^bQ_{\rho}^c\prod_{n>0,l\in\mathbb{Z}}(1-Q_{\tau}^nQ_{m}^l)^{c(0,l)}\prod_{n=m=0,l<0}(1-Q_{m}^l)^{c(0,l)}\nonumber\\&&=
Q_{\rho}^cQ_{\tau}^{\frac{1}{24}c(0,0)}\prod_{n>0}(1-Q_{\tau}^n)^{c(0,0)}\prod_{l>0}(Q_{m}^{\frac{l}{2}}-Q_{m}^{\frac{-l}{2}})^{c(0,l)}Q_{\tau}^{\frac{1}{12}\sum_{l>0}c(0,l)}\prod_{n>0,l>0}\bigg((1-Q_{\tau}^nQ_{m}^l)(1-Q_{\tau}^nQ_{m}^{-l})\bigg)^{c(0,l)}\nonumber\\&&=
Q_{\rho}^cQ_{\tau}^{\frac{1}{24}c(0,0)}\prod_{n>0}(1-Q_{\tau}^n)^{c(0,0)}\prod_{n>0,l>0}\frac{\bigg(Q_{\tau}^{\frac{1}{8}}(Q_{m}^{\frac{l}{2}}-Q_{m}^{\frac{-l}{2}})(1-Q_{\tau}^nQ_{m}^l)(1-Q_{\tau}^nQ_{m}^{-l})(1-Q_{\tau}^n)\bigg)^{c(0,l)}}{Q_{\tau}^{\frac{c(0,l)}{24}}(1-Q_{\tau}^n)^{c(0,l)}}\nonumber\\&&=Q_{\rho}^c\eta(\tau)^{c(0,0)}\prod_{l>0}\frac{\theta_1(\tau;lz)^{c(0,l)}}{\eta(\tau)^{c(0,l)}}
\eea

Therefore we can write the Borcherds product as
\bea
\Phi_k=Q_{\rho}^c\eta(\tau)^{c(0,0)}\prod_{l>0}\frac{\theta_1(\tau;lz)^{c(0,l)}}{\eta(\tau)^{c(0,l)}}e^{L\phi_{0,1}^k}
\eea
\section{Determinant bundles and Ray-Singer Torsion}\label{sec:DRS}
Consider a proper smooth morphism (\cite{1999math......4092Y}) $\pi:\mbox{X}\to \mbox{S}$ of K\"ahler manifolds X and S and a determinant line bundle $\lambda(\mathcal{O}_X)$ on X defined as \cite{1999math......4092Y}:
\bea
\lambda(\mathcal{O}_X):=\otimes_{q\ge 0}(\mbox{det}\mbox{R}^q\pi_{*}\mathcal{O}_X)^{(-1)^q}
\eea 
There exists a Hermitian metric $g_{X/S}$ on $TX/S:=\mbox{mer}\pi_*$ whose restriction to any fiber $X_t=\pi^{-1}(t)$ is K\"ahler. The line bundle $\lambda(\mathcal{O}_X)_t$ can thus be expressed as
\bea
\lambda(\mathcal{O}_X)_t:=\otimes_{q\ge 0}(\wedge^{max}\mathcal{H}^{0,q}(X_t))^{(-1)^q}
\eea
for the space of $(0,q)$-forms $\mathcal{H}^{0,q}$. This identification of determinant line bundle endows it with an $L^2$-metric relative to $g_{X/S}$ denoted by $||.||_{L^2}$. The norm $||.||^2_{L^2}(t)$ is not smooth but can be made so by multiplying it by holomorphic torsion $I(X_t)$ of the fiber $X_t$. The modified smooth metric is called Quillen metric $||.||^2_Q(t)$ of $\lambda(\mathcal{O}_{X_t})$ and expressed as
\bea
||.||^2_Q(t):=I(X_t)||.||^2_{L^2}(t)
\eea
The Quillen  anomaly has its origin in the non-triviality of the line bundle $\lambda(\mathcal{O}_X)_t$ . 
The symplectic group $\Gamma_g=Sp(2g,\mathbb{Z})$ acts on the theta divisors and is uplifted to its action on the corresponding line bundles $\mbox{L}$. However only the subgroup $\Gamma_g(1,2):=\{ \begin{pmatrix}A &B\\ C&D \end{pmatrix} \}\in\Gamma_g'(A^tC)_0\equiv (B^tD)_0\equiv 0\quad \mbox{mod}\quad 2$
where $X_0=x_{ij}\delta_{ij}$ is  the diagonal of $X=x_{ij}$, preserves the line bundle. 
The space of sections $s_c\in H^0(\mathbb{P}(V_m)\times \mathbb{H}_g,\lambda(\mathcal{O}_{\Theta_m})^{(-1)^g})$  
is generated by \cite{1999math......4092Y}
\bea
s_c:=\Large\wedge_{a\in B_m}\frac{u_c\theta_a}{\sum_{b\in B_m}u_b\theta_b}dz_1\wedge dz_2\wedge...\wedge dz_g
\eea
In fact for $g>1$ and $m\ge 2$ the space of functions $\{\sigma_J\}_{|J|=m^g}$ defined as
\bea
\sigma_J|_{\mathcal{U}_c\times\mathbb{H}_g}:=\frac{u^J}{u_c^{m^g}}s_c
\eea
gives a map to
 a well defined basis of 
$ \lambda(\mathcal{O}_{\Theta_m})^{(-1)^g}$. For $m=1$ one gets an expression for  section of $\lambda_{\mathcal{O}_\Theta}$ as
 \bea
 \sigma_{\mathcal{O}_\Theta}:=1_{\mathbb{A}}\otimes(dz_1\wedge\wedge...\wedge dz_g)^{(-1)^g}
\eea
The Quillen norm of $\sigma_J$ is given by
\bea
||\sigma_J||^2_Q=(\mbox{detIm}\tau)^{\frac{(g-1)m^g}{2(g+1)}}|\frac{u^J}{\Delta_{g,m}(u,\tau)^{\frac{1}{(g+1)!}}}|^2
\eea
where $\Delta_{g,m}(u,\tau)^{\frac{1}{(g+1)!}}$ is a holomorphic function homogenous in variable $u_i$ and satisfies the following automorphy property
\bea
\Delta_{g,m}(\gamma.u,\gamma.\tau)=\gamma_{g,m}(\gamma,u,\tau)j(\tau,\gamma)^{\frac{(g+3)g!m^g}{2}}\Delta_{g,m}(u,\tau)
\eea
Moreover $\Delta_{g,m}(u,\tau)$ defines the projective surface $\mathcal{D}_{g,m,\tau}$ which is the fiber of the map $p:\mathcal{D}_{g,m}\to \mathbb{H}_g$.\\
We can specialise to $m=1$ corresponding to the the smooth theta divisor $\Theta$  and find the Quillen norm as
\bea
||\sigma_{\Theta}||^2_Q=(\mbox{detIm}\tau)^{\frac{(-1)^g(g-1)}{2(g+1)}}|\Delta_{g}(\tau)^{\frac{(-1)^{g+1}}{(g+1)!}}|^2
\eea
On the other hand the $L^2$-norm can be calculated to be
\bea
||\sigma_{\Theta}||^2_{L^2(\Theta)}=(\mbox{det2Im}\tau)^{(-1)^g}
\eea
 The ratio $\frac{||\sigma_{\Theta}||^2_Q}{||\sigma_{\Theta}||^2_{L^2(\Theta)}}$ defines the Ray-Singer torsion $I$.
\\
The embedding or projective compactification corresponding to the complete linear system $|L_{m,n,\tau}|$ is defined by the theta variables $\theta_{a,b}$ with characteristics $a\in\frac{1}{m}\mathbb{Z}^g,b\in\frac{1}{n}\mathbb{Z}^g$. The theta coordinates $\theta_{a,b}$ define the sections of line bundle $L_{m,n}$ and also the embedding in the projective space $\mathbb{P}^{m^gn^g-1}$ by the algebraic relations for $m,n\ge 4$ \cite{HokPunYu:2009hpy}
\bea\label{eq:embedding}
\theta_{y_1,\rho}\sum_{z\in Z_2}\rho(z)X_{y+y_2+z}X_{y-y_2+z}=\theta_{y_2,\rho}(0)\sum_{z\in Z_2}\rho(z)X_{y+y_1+z}X_{y-y_1+z}
\eea
where $\rho(z)$ are the characters of $\mbox{Hom}(X_2,\pm1)$ and
\bea
\theta_{y_i,\rho}(0)&=&\sum_{z\in Z_2}\rho(z)\theta_{y_i+z}(0)^2,\nonumber\\
\theta_y:&=&\theta\begin{pmatrix}a\\ b \end{pmatrix}(z)
\eea
for $y,y_1,y_2\in \mathbb{Z}_m\oplus \mathbb{Z}_n$ and $y\equiv y_1\equiv y_2 \quad \mbox{mod}\quad2\mathbb{Z}_m\oplus2\mathbb{Z}_n$. The eq.(\ref{eq:embedding}) describes the projectification of the linear system $|L_{m,\tau}|$ in terms of the homogeneous coordinates $X_a$ for $a\in B_m$.
As indicated before $\mbox{Divisor}(\Delta_{g,m})=\mathcal{D}_{g,m}$. For a given $J$ we can expand $\Delta_{g,m}(u,\tau)$ in terms of holomorphic functions $f_J(\tau)$ as follows
\bea
\Delta_{g,m}(u,\tau)=\sum_Jf_J(\tau)u^J
\eea
and it can be shown that
\bea
f_{J_0}(\tau)=f_{(m^g(g+1)!,0,...,0)}(\tau)=m^{\frac{g g!m^g}{2}}\Delta_{g}(\tau)^{m^g}
\eea
To elucidate the structure we can consider the normalised modular form $\frac{\Delta_{g,m}(u,\tau)}{m^{\frac{g g!m^g}{2}}\Delta_{g}(\tau)^{m^g}}$
\bea
\tilde{\Delta}_{g,m}(u,\tau):&=&\frac{\Delta_{g,m}(u,\tau)}{m^{\frac{g g!m^g}{2}}\Delta_{g}(\tau)^{m^g}}\nonumber\\&=&u_0^{m^g}+\sum_{J\ne J_0}\frac{f_J(\tau)}{f_{J_0}(\tau)}u^J
\eea
 The algebraic equations (\ref{eq:embedding}) carve an algebraic variety $\mathcal{A}_m$ in $\mathbb{P}_k^{m^g}$. The projective dual variety $\check{\mathcal{A}}_m$ is described by the equation of  the modular form $\tilde{\Delta}_{g,m}(u,\tau)$. Notice that this modular from is homogeneous in variables $u_i$ and monic in $u_0$. The modular form $\tilde{\Delta}_{g,m}(u,\tau)$ can be expressed in terms of theta constants i.e. $\tilde{\Delta}_{g,m}(u,\tau)\in \mathbb{Q}[\theta_{a,0}(0,m\tau)\theta_{b,0}(0,m\tau)]_{a,b\in B_m}[u_a]_{a\in B_m}$,
 moreover there exists a constant $C_g$ such that  the ring $\mathbb{Z}[\theta_{a,b}(0,\tau)\theta_{c,d}(0,\tau)]_{a,b,c,d\in B_2}$ contains $C_g^{-1}\Delta_g(\tau) $ with the Fourier coefficients of the later taking values in $\mathbb{Q}$.


\bibliographystyle{plain}
\bibliography{bibliography}

\begin{thebibliography}{10}

\bibitem{Aharony:1999ks}
Ofer Aharony.
\newblock {A Brief review of 'little string theories'}.
\newblock {\em Class. Quant. Grav.}, 17:929--938, 2000.

\bibitem{Aoki2005SIMPLEGR}
Hiroki Aoki and T.~Ibukiyama.
\newblock Simple graded rings of siegel modular forms, differential operators and borcherds products.
\newblock {\em International Journal of Mathematics}, 16:249--279, 2005.

\bibitem{Belin:2016knb}
Alexandre Belin, Alejandra Castro, Joao Gomes, and Christoph~A. Keller.
\newblock {Siegel Modular Forms and Black Hole Entropy}.
\newblock {\em JHEP}, 04:057, 2017.

\bibitem{Belin:2018oza}
Alexandre Belin, Alejandra Castro, Joao Gomes, and Christoph~A. Keller.
\newblock {Siegel paramodular forms and sparseness in AdS$_{3}$/CFT$_{2}$}.
\newblock {\em JHEP}, 11:037, 2018.

\bibitem{Belin:2019jqz}
Alexandre Belin, Alejandra Castro, Christoph~A. Keller, and Beatrix~J. M\"uhlmann.
\newblock {Siegel Paramodular Forms from Exponential Lifts: Slow versus Fast Growth}.
\newblock 10 2019.

\bibitem{Bershadsky:1993ta}
M.~Bershadsky, S.~Cecotti, H.~Ooguri, and C.~Vafa.
\newblock {Holomorphic anomalies in topological field theories}.
\newblock {\em Nucl. Phys. B}, 405:279--304, 1993.

\bibitem{Bershadsky:1993cx}
M.~Bershadsky, S.~Cecotti, H.~Ooguri, and C.~Vafa.
\newblock {Kodaira-Spencer theory of gravity and exact results for quantum string amplitudes}.
\newblock {\em Commun. Math. Phys.}, 165:311--428, 1994.

\bibitem{Borcherds1995}
Richard~E. Borcherds.
\newblock Automorphic forms on os + 2,2(r) and infinte products.
\newblock {\em Inventiones mathematicae}, 120(1):161--214, 1995.

\bibitem{Borcherds:1996uda}
Richard~E. Borcherds.
\newblock {Automorphic forms with singularities on Grassmannians}.
\newblock {\em Invent. Math.}, 132:491--562, 1998.

\bibitem{2019arXiv190701535B}
Jim {Bryan} and {\'A}d{\'a}m {Gyenge}.
\newblock {$G$-fixed Hilbert schemes on $K3$ surfaces, modular forms, and eta products}.
\newblock {\em arXiv e-prints}, page arXiv:1907.01535, July 2019.

\bibitem{Cecotti:1992vy}
Sergio Cecotti and Cumrun Vafa.
\newblock {Ising model and N=2 supersymmetric theories}.
\newblock {\em Commun. Math. Phys.}, 157:139--178, 1993.

\bibitem{Dabholkar:2012nd}
Atish Dabholkar, Sameer Murthy, and Don Zagier.
\newblock {Quantum Black Holes, Wall Crossing, and Mock Modular Forms}.
\newblock 8 2012.

\bibitem{Dabholkar:2006xa}
Atish Dabholkar and Suresh Nampuri.
\newblock {Spectrum of dyons and black holes in CHL orbifolds using Borcherds lift}.
\newblock {\em JHEP}, 11:077, 2007.

\bibitem{Dijkgraaf:2007sw}
Robbert Dijkgraaf, Lotte Hollands, Piotr Sulkowski, and Cumrun Vafa.
\newblock {Supersymmetric gauge theories, intersecting branes and free fermions}.
\newblock {\em JHEP}, 02:106, 2008.

\bibitem{Dijkgraaf:1996xw}
Robbert Dijkgraaf, Gregory~W. Moore, Erik~P. Verlinde, and Herman~L. Verlinde.
\newblock {Elliptic genera of symmetric products and second quantized strings}.
\newblock {\em Commun. Math. Phys.}, 185:197--209, 1997.

\bibitem{Dijkgraaf:2002yn}
Robbert Dijkgraaf, Annamaria Sinkovics, and Mine Temurhan.
\newblock {Matrix models and gravitational corrections}.
\newblock {\em Adv. Theor. Math. Phys.}, 7(6):1155--1174, 2003.

\bibitem{Dijkgraaf:1996it}
Robbert Dijkgraaf, Erik~P. Verlinde, and Herman~L. Verlinde.
\newblock {Counting dyons in N=4 string theory}.
\newblock {\em Nucl. Phys. B}, 484:543--561, 1997.

\bibitem{Dijkgraaf:2002ac}
Robbert Dijkgraaf, Erik~P. Verlinde, and Marcel Vonk.
\newblock {On the partition sum of the NS five-brane}.
\newblock 5 2002.

\bibitem{Eguchi:2011aj}
Tohru Eguchi and Kazuhiro Hikami.
\newblock {Twisted Elliptic Genus for K3 and Borcherds Product}.
\newblock {\em Lett. Math. Phys.}, 102:203--222, 2012.

\bibitem{eichler2013theory}
M.~Eichler and D.~Zagier.
\newblock {\em The Theory of Jacobi Forms}.
\newblock Progress in Mathematics. Birkh{\"a}user Boston, 2013.

\bibitem{Gopakumar:1998ki}
Rajesh Gopakumar and Cumrun Vafa.
\newblock {On the gauge theory / geometry correspondence}.
\newblock {\em Adv. Theor. Math. Phys.}, 3:1415--1443, 1999.

\bibitem{Govindarajan:2010fu}
Suresh Govindarajan.
\newblock {BKM Lie superalgebras from counting twisted CHL dyons}.
\newblock {\em JHEP}, 05:089, 2011.

\bibitem{Griffiths:433962}
Phillip~A Griffiths and Joseph Harris.
\newblock {\em {Principles of algebraic geometry}}.
\newblock Wiley classics library. Wiley, New York, NY, 1994.

\bibitem{Gritsenko:1999nm}
V.~Gritsenko.
\newblock {Complex vector bundles and Jacobi forms}.
\newblock 1 1999.

\bibitem{Gritsenko:1996tm}
Valeri~A. Gritsenko and Vyacheslav~V. Nikulin.
\newblock {Automorphic forms and Lorentzian Kac-Moody algebras. Part 2}.
\newblock 12 1996.

\bibitem{Harvey:1996ts}
Jeffrey~A. Harvey and Gregory~W. Moore.
\newblock {Exact gravitational threshold correction in the FHSV model}.
\newblock {\em Phys. Rev. D}, 57:2329--2336, 1998.

\bibitem{Hohenegger:2013ala}
Stefan Hohenegger and Amer Iqbal.
\newblock {M-strings, elliptic genera and $\mathcal{N} = 4$ string amplitudes}.
\newblock {\em Fortsch. Phys.}, 62:155--206, 2014.

\bibitem{Hohenegger:2015cba}
Stefan Hohenegger, Amer Iqbal, and Soo-Jong Rey.
\newblock {M-strings, monopole strings, and modular forms}.
\newblock {\em Phys. Rev. D}, 92(6):066005, 2015.

\bibitem{Hohenegger:2015btj}
Stefan Hohenegger, Amer Iqbal, and Soo-Jong Rey.
\newblock {Instanton-monopole correspondence from M-branes on $\mathbb S^1$ and little string theory}.
\newblock {\em Phys. Rev. D}, 93(6):066016, 2016.

\bibitem{Hohenegger:2016eqy}
Stefan Hohenegger, Amer Iqbal, and Soo-Jong Rey.
\newblock {Self-Duality and Self-Similarity of Little String Orbifolds}.
\newblock {\em Phys. Rev. D}, 94(4):046006, 2016.

\bibitem{Hollowood:2003cv}
Timothy~J. Hollowood, Amer Iqbal, and Cumrun Vafa.
\newblock {Matrix models, geometric engineering and elliptic genera}.
\newblock {\em JHEP}, 03:069, 2008.

\bibitem{Iqbal:2007ii}
Amer Iqbal, Can Kozcaz, and Cumrun Vafa.
\newblock {The Refined topological vertex}.
\newblock {\em JHEP}, 10:069, 2009.

\bibitem{Kawai:2000px}
Toshiya Kawai and Kota Yoshioka.
\newblock {String partition functions and infinite products}.
\newblock {\em Adv. Theor. Math. Phys.}, 4:397--485, 2000.

\bibitem{Minahan:1998vr}
J.~A. Minahan, D.~Nemeschansky, C.~Vafa, and N.~P. Warner.
\newblock {E strings and N=4 topological Yang-Mills theories}.
\newblock {\em Nucl. Phys. B}, 527:581--623, 1998.

\bibitem{Vafa:1994tf}
Cumrun Vafa and Edward Witten.
\newblock {A Strong coupling test of S duality}.
\newblock {\em Nucl. Phys. B}, 431:3--77, 1994.

\bibitem{HokPunYu:2009hpy}
Xiaowei Wang and Hok~Pun Yu.
\newblock {Theta function and Bergman metric on Abelian varieties}.

\bibitem{1999math......4092Y}
Ken-Ichi {Yoshikawa}.
\newblock {Discriminant of theta divisors and Quillen metrics}.
\newblock {\em arXiv Mathematics e-prints}, page math/9904092, April 1999.

\end{thebibliography}

\end{document}